\newcommand{\caproman}[1]{\uppercase\expandafter{\romannumeral#1}}
\newcommand{\bs}{$\backslash$}
\newcommand{\BGsite}{http://physics.harvard.edu/\raisebox{-4pt}{$\tilde{\;\;}$}gottschalk}
\newcommand{\BGmail}{bgottsch\,@\,fas.harvard.edu}

\documentclass{article}
\usepackage{amsmath}
\usepackage{latexsym}
\usepackage{graphicx}

\begin{document}

\title{\bf On the scattering power of radiotherapy protons}
\author{Bernard Gottschalk
\thanks{\;\;Harvard University Laboratory for Particle Physics and Cosmology, 18 Hammond St., Cambridge, MA 01238, USA,\;\BGmail}}
\maketitle

\begin{abstract}
\noindent
Scattering power $T\equiv d<\theta^2>/dx$, used in proton transport calculations, is properly viewed as a differential description of the Gaussian approximation to multiple Coulomb scattering theories such as that of Moli\`ere. Accurate formulas for $T$ must take into account the competition between the Gaussian core and the single scattering tail of the angular distribution, which affects the rate of change of the Gaussian width, and leads to the {\em single scattering correction}. Mathematically, that implies that $T$ must be {\em nonlocal}. In addition to proton energy and properties of the scattering material at the point of interest, it must depend in some way on how far the multiple scattering process has proceeded.

We review five previous formulas for $T$ and propose a sixth, `differential Moli\`ere', formula
\[T_\mathrm{dM}\;\equiv\;f_{dM}(pv,p_1v_1)\times\left(\frac{E_s}{pv}\right)^2\;\frac{1}{X_S}\]
where
\[f_\mathrm{dM}\;\equiv\;0.5244+0.1975\lg(1-(pv/p_1v_1)^2)+0.2320\lg(pv)-0.0098\lg(pv)\lg(1-(pv/p_1v_1)^2)\]
is a fit to the single scattering correction as deduced from Moli\`ere/Fano/Hanson theory. The scattering length $X_S$ is a new material property similar to radiation length. $pv$ is the product of proton momentum and speed at the point of interest, $p_1v_1$ the same at the initial energy, and $E_s=15.0$\;Mev. $T_\mathrm{dM}$ is easily computed and generalizes readily to mixed slabs because $f_\mathrm{dM}$ is not explicitly material dependent.

Whether or not an accurate formula for $T$ is required depends very much on the problem at hand. For beam spreading in water, five of the six formulas for $T$ give almost identical results, suggesting that patient dose calculations are insensitive to $T$. That is not as true of beam spreading in Pb. Evidently some favorable cancelation occurs in low-$Z$ materials. At the opposite extreme, the projected rms beam width at the end of a Pb/Lexan/air stack, analogous to the upstream modulator in a passive beam spreading system, is sensitive to the choice of $T$. A simple experiment would discriminate between all but two of the six formulas.

The usefulness of scattering power as a concept applies just as much to Monte Carlo as to deterministic transport calculations. For instance, using $T$ in any of its forms will avoid step size dependence. Using the best available $T$ might be important in general purpose Monte Carlo codes, which are expected to give the correct answer to many different problems.
\end{abstract}
\clearpage


\section{Introduction}
Consider an initially monoenergetic, monodirectional proton beam slowing down in a finite slab of some homogeneous material. At any depth $x$ we can write down the stopping power $S\equiv -dE/dx$, the rate of decrease of energy with depth. There exists a well-known theory \cite{icru49} for computing $S$, which depends only on the mean proton energy at $x$ and atomic properties of the stopping medium. Integrating $S$ over $x$ we obtain, accurately, the total energy loss in a finite slab.

At any $x$ we can also calculate a scattering power
\begin{equation}T\equiv d<\theta^2>/dx\label{eqn:T}\end{equation}
the rate of increase with $x$ of the mean squared projected multiple Coulomb scattering (MCS) angle.\footnote{~In the early literature $T$ is the rate of increase of the mean squared {\em space} angle, which is greater by a factor 2. In transport calculations projected quantities are usually more convenient.} There are a number of formulas for $T$ in the literature. According to some of these $T$, too, depends only on local variables at $x$ (mean proton energy and atomic properties of the stopping medium). However if we integrate these over $x$ we do not, in general, obtain the correct rms MCS angle. The analogy with $S$ is flawed. 

This might seem irrelevant since there exist accurate and well tested theories, notably that of Moli\`ere \cite{moliere2,bethe,mcsbg}, which directly give the total MCS angle, at arbitrary incident energy, for homogeneous slabs of any element, compound or mixture from very thin to near stopping thickness. Moreover, the derivation \cite{moliere2,bethe} of Moli\`ere theory does not rely on the concept `stopping power'. It is never mentioned.

Seen in that light, the aim of this paper is to discuss differential approximations to Moli\`ere theory, functions $T$ which, if integrated over $x$ for any slab, will recover the Moli\`ere MCS angle more or less accurately.

Why do we want to do that? Because practical problems in proton transport go far beyond merely finding the MCS angle in a finite slab. We might for instance wish to compute transverse spreading of a pencil beam in a single slab, or in a stack of slabs, or the equivalent source point\footnote{~We use the term loosely here. Various `equivalent sources' are defined in the literature cited below.} in a stack, or the transverse penumbra of a proton therapy beam formed by such a stack. Such computations are done either by Monte Carlo methods or by deterministic (numerical) methods using the Gaussian approximation to proton transport, generalized Fermi-Eyges theory.\footnote{~We assume some familiarity with generalized Fermi-Eyges theory. For reviews see, for instance, ICRU Report\;35 \cite{icru35} or more recent papers by Hollmark et al. \cite{hollmark} and Kanematsu \cite{kanematsu08}.} Either way, we need to know, at every depth, the rate of change of the mean squared MCS angle. That need is exemplified by the formulas
\begin{equation}\label{eqn:An}
A_n(x)\;\equiv\;\int_0^x(x-x')^n\,T(x')\,dx'\quad,\quad n=0,1,2
\end{equation}
for the Fermi-Eyges moments, but it is more fundamental than that. Any transport calculation ultimately needs a recipe for the local rate of change of whatever variables are involved.

It should go without saying that, when we do integrals like (\ref{eqn:An}) numerically, approximating them by sums as in the midpoint rule or Simpson's rule, the answer should not depend on the step size $\Delta x$ over some reasonable range. The same applies if $\Delta x$ happens to be the step size in a Monte Carlo calculation. We mention this only because, in the case of $T$, this issue has lead to some confusion. For instance, Li and Rogers \cite{li} in an electron paper\footnote{~For historical reasons Fermi-Eyges theory and the concept of $T$ were first developed (in the radiotherapy context) for electrons, but the conceptual issues are the same for protons. Indeed, protons are much better Fermi-Eyges particles. The small angle approximation comes naturally and the correlation of mean energy with depth holds to far greater depth because range straggling and the MCS detour factor are much smaller.} discuss the `slab thickness dependence' of $T$ at length, while a proton paper by Russell et al. \cite{russell} uses, without much discussion, Moli\`ere theory with the step size (there conflated with slab thickness) fixed at an arbitrary value.

We shall see that there is, in fact, no problem provided we keep the concepts `step size', `slab thickness' and `depth of the point of interest (POI)' carefully separated. All numerical results given later hold over a wide range of step size. We often use $\Delta x$\;(cm) equivalent to 0.1\;g/cm$^2$ but a much smaller or larger step gives essentially the same answer.

The present paper deals with $T$ and its implications in {\em mixed slab} geometry, with the slabs assumed to be infinite transversely. In most practical problems, whether beam line design or dose distribution in the patient, transverse heterogeneity or beam limiting devices eventually come into play. We must go beyond Fermi-Eyges theory, which is then relegated to propagating the individual pencil beams which make up the final (non Gaussian) distributions. The present paper ignores all that. We treat only the essentially one dimensional mixed slab problem, in order to understand that fully before introducing complications. Nor do we consider non electromagnetic effects like the halo from non-elastic nuclear reaction secondaries \cite{pedroniPencil,hollmark08}. This is strictly about the Gaussian approximation to MCS in mixed slab geometry.

This paper overlaps a recent one by Kanematsu \cite{kanematsu08}, whose notation we use except as noted. That work discusses the entire transport problem in the framework of Fermi-Eyges theory, treats heavy ions as well as protons, introduces a new scattering power $T_\mathrm{dH}$ and presents many useful analytic approximations. The present paper focuses entirely on scattering power, protons, and single or mixed slabs, largely glossing over computational issues. It is an in-depth look at a small part of the ground covered by Kanematsu. His work is somewhat biased towards dose reconstruction in the patient, ours towards beam line design, but the fundamental issues are the same and there is no significant disagreement.

For completeness, we repeat a number of formulas from the literature, some of which is quite old. There is, however, considerable new content: simplifying the formula we call $T_\mathrm{IC}$, deriving a new non-local formula $T_\mathrm{dM}$, and proposing, for a case of practical importance, an experimental test  to discriminate between various formulas for $T$. In short, we combine original material with a critical review which, we hope, will be useful to the student and of some interest to the expert.

\section{Preliminaries}

\subsection{Mixed-Slab Notation}\label{sec:mixed}

\begin{figure}[p]
\centering\includegraphics[width=4.87in,height=3.5in]{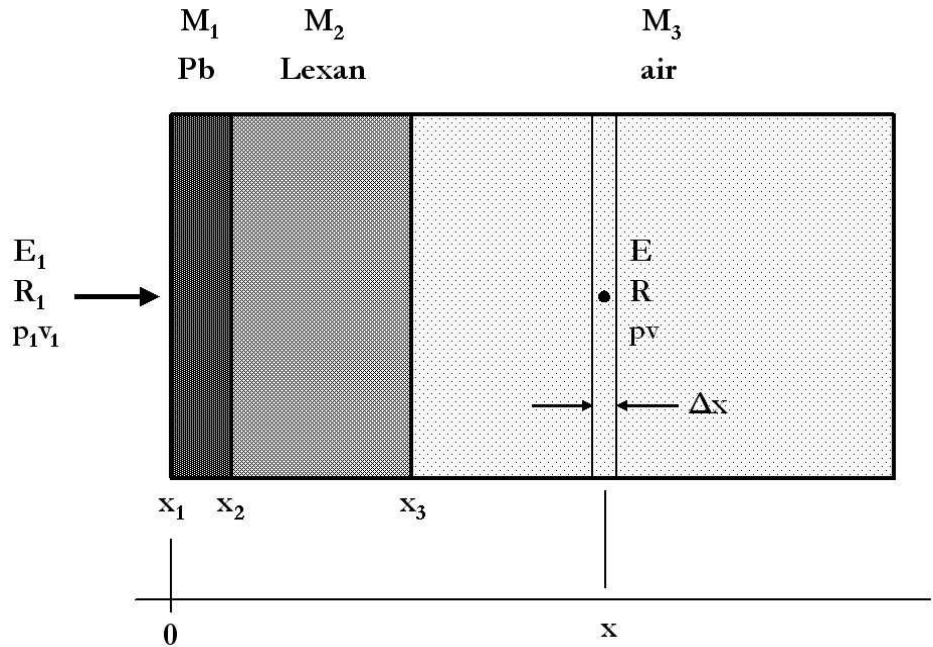} 
\caption{Mixed slab (stack) geometry.\label{fig:Stack3}}
\end{figure}

Take, for concreteness, an example from beam line design: a monoenergetic proton beam of known emittance\footnote{~In other words the parameters of the incident Fermi-Eyges beam ellipse \cite{icru35} are known to sufficient accuracy.} enters a Pb/Lexan/air stack (Figure \ref{fig:Stack3}). That is precisely what happens at the upstream range modulator/first scatterer of a modern passive beam spreading system \cite{delaney}.\footnote{~An `upstream' modulator also serves as the first scatterer in a double scattering system, and MCS in it is critical. A `downstream' modulator is near the patient and MCS has little effect.} (In the patient, the variation of material properties is usually less extreme, but still present.) We wish to transport the beam through the stack to an arbitrary measuring plane MP at a depth $x$ (cm) measured from the entrance of the first slab. In particular let us assume we wish to find $y_{rms}$ of the Gaussian which approximately describes the transverse fluence distribution on an MP located, say, at the second scatterer, because the precise match of $y_{rms}$ to the second scatterer design \cite{grusell} will determine the flatness of the dose distribution at the patient.

$M_i$ is the material of which the $i^\mathrm{th}$ slab is composed and stands for a set of material properties e.g. density $\rho_i$ (g/cm$^3$), radiation length $X_{0i}$ (cm) and (to be defined) scattering length $X_{Si}$ (cm). There are also atomic weights $A$, atomic numbers $Z$ and fractions by weight $w$ of the constituents of $M_i$, all implicit stepwise constant functions of $x$. $x_i$ refers to the entrance of the $i^\mathrm{th}$ slab and $x_1\equiv0$. 

$E_i$ (MeV) is the proton kinetic energy entering the $i^\mathrm{th}$ slab, $p_iv_i$ (MeV) the corresponding product of momentum and speed, and $R_i$ (cm) the mean proton range corresponding to $E_i$ stopping in $M_i$, that is to say the residual range in $M_i$ at the entrance to the $i^\mathrm{th}$ slab. $E$, $pv$ and $R$ (the latter always computed for the current material) are corresponding quantities at the depth of interest $x$. Given a specified stack and any value of $x$ within it, all these quantities can be computed using range-energy tables and kinematic relations.\footnote{~In the early literature $x$ is frequently expressed in units of the radiation length $X_0$ (therefore dimensionless) and instead of $T$, the mass stopping power $T/\rho$ (radian$^2$/(g/cm$^2$)) is used. This is unhelpful for mixed slabs.}

This example is a rare practical case where Fermi-Eyges theory alone gives a useful answer, because there is in fact no transverse heterogeneity and the final fluence distribution is in fact very nearly Gaussian. In more complicated beam line design problems, with collimators as well as scatterers whose thickness varies with radius, Fermi-Eyges theory can only be a building block, as noted earlier.

\subsection{Limits of Moli\`ere Theory}\label{sec:limits}
Radiotherapy protons have kinetic energies roughly in the range $3\le E\le 300$\,MeV corresponding to $0.015\le R_1\le51.5$\,cm in water. The interesting range of normalized depth in a single slab is roughly $0.001\le x/R_1\le 0.97$. The lower limit might for instance apply to  a vacuum window in a beam scanning system. The upper, near stopping depth, is where Moli\`ere theory fails \cite{mcsbg} as straggling blurs the correlation between depth and energy. It is reached and exceeded when protons stop in the patient.

It is well known that Moli\`ere theory applies only to slabs of some minimum thickness. There must be enough atomic encounters to establish  the regime of multiple (as distinct from single or plural) scattering.\footnote{~One can speak of the (extremely small) average energy loss in an atomic monolayer, or of the (extremely small) single scattering probability, but it makes no sense at all to speak of multiple scattering since there is at most one collision. That is the fundamental difference between stopping and multiple scattering theories.} Moli\`ere \cite{moliere2} gives $B=4.5$ as the lowest allowed value of his slab thickness parameter.\footnote{~$B\propto$\;logarithm of normalized slab thickness. The constant of proportionality varies with the material \cite{mcsbg}.} Table \ref{tbl:B} lists target parameters at \begin{table}[h]
\begin{center}\begin{tabular}{crrcrr}
&\multicolumn{1}{c}{$\rho$}&\multicolumn{1}{c}{$\rho\,R_1$}&\multicolumn{1}{c}{$x$}&
\multicolumn{1}{c}{$x/R_1$}&\multicolumn{1}{c}{$\theta_\mathrm{Hanson}$}\\ \vspace{3pt}
&\multicolumn{1}{c}{g/cm$^3$}&\multicolumn{1}{c}{g/cm$^2$}&\multicolumn{1}{c}{$\mu$\,m}&
\multicolumn{1}{c}{$\times10^6$}&\multicolumn{1}{c}{mrad}\\
Be&1.85&21.29&0.65&5.6&0.025\\
water&1.00&17.38&2.35&13.5&0.051\\
air&0.0012&19.67&0.24\,cm&14.4&0.056\\
Cu&8.96&26.26&0.79&27.0&0.167\\
Pb&11.35&36.06&2.11&66.6&0.483
\end{tabular}\end{center}
\caption{Target parameters for which Moli\`ere's $B$ equals 4.5 at 158.6\,MeV incident.\label{tbl:B}}
\end{table}
which that limit is reached for some common materials for 158.6\,MeV incident protons. It is clear that the lower Moli\`ere limit need not concern us in practical proton radiotherapy calculations.

\subsection{Kinematics}\label{sec:kinematics}
The kinematic expression $1/pv$ appears in all multiple scattering formulas.\footnote{~Multiple scattering is derived from single scattering as outlined in Section\,\ref{sec:TFR}. $1/v$ enters  the derivation of the single scattering probability (\ref{eqn:Xi}) because the impulse delivered in a single collision is proportional to the interaction time or inverse speed. $1/p$ comes from the fact that the angle of deflection equals $\Delta p/p$.} If we define the reduced kinetic energy of any particle as
\begin{equation}\tau\;\equiv\;\frac{E}{mc^2}\label{eqn:tau}\end{equation}
where $mc^2$ is the particle's rest energy, we find
\begin{equation}pv\;=\;\frac{\tau+2}{\tau+1}\;E\end{equation}
For $3\le E\le 300$\,MeV protons the coefficient of $E$ varies from 2.00 to 1.76. Other useful relations in the same vein are
\begin{equation}\beta^2\;=\;\frac{\tau+2}{(\tau+1)^2}\;\tau\label{eqn:betasq}\end{equation}
and
\begin{equation}(pc)^2\;=\;(\tau+2)\,mc^2\,E\label{eqn:pcsq}\end{equation}
These formulas avoid differences between large quantities that can arise in relativistic calculations, and their relativistic and non-relativistic limits are obvious by inspection.

\subsection{The {\O}ver{\aa}s Approximation}\label{sec:Overas}
\begin{figure}[p]
\centering\includegraphics[width=4.72in,height=3.5in]{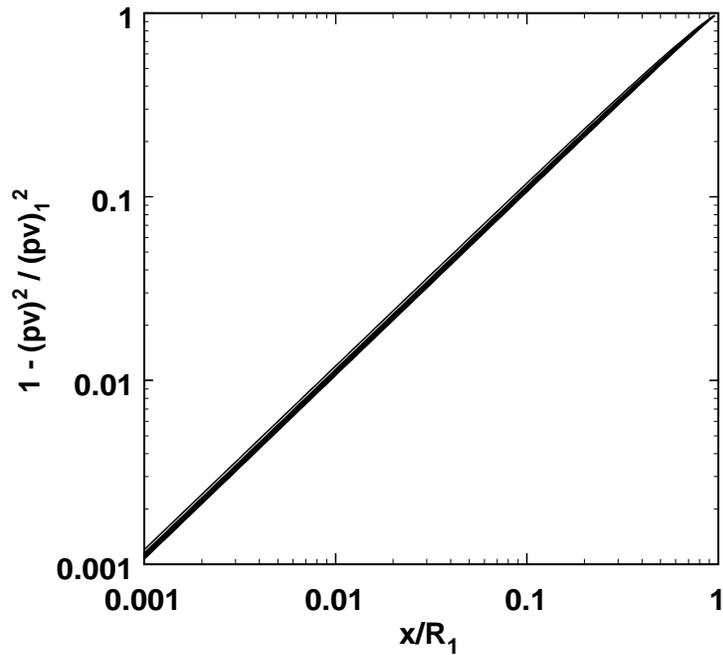} 
\caption{Weak {\O}ver{\aa}s approximation: the function $1-(pv)^2/(pv)_1\,^2$ as a function of normalized slab thickness for protons of energies 20, 50, 100 and 200\;MeV exiting single slabs of Be, Cu and Pb (all superimposed). 20\;MeV protons leaving a very thin Pb slab ($x/R_1=0.001$) have the largest deviation from the ideal, 20\%. \label{fig:PVfuncVsTnLL}}
\end{figure}

{\O}ver{\aa}s \cite{overas} found a simple empirical relation between $(pv)^2$ and normalized residual range. For a single slab 
\begin{equation}\label{eqn:overas}
(pv)^2\;=\;(p_1v_1)^2\;(1-x/R_1)^{(1+k)}
\end{equation}
where $k$ depends on the material. If the material is characterized by its radiation length, Schneider et al. \cite{schneider01} found the empirical expression 
\begin{equation}\label{eqn:k}
k=0.12\;e^{\textstyle-0.09\,\rho X_0}\;+\;0.0753
\end{equation}
$k$ is fairly small compared to 1, and often the `weak {\O}ver{\aa}s' approximation
\begin{equation}\label{eqn:overasWeak}
(pv)^2\;\approx\;(p_1v_1)^2\;(1-x/R_1)
\end{equation}
is adequate. We shall use it later as a guideline. Figure \ref{fig:PVfuncVsTnLL} tests (\ref{eqn:overasWeak}) over a wide range of materials, normalized slab thicknesses and outgoing kinetic energies. 

\subsection{Moli\`ere/Fano/Hanson Procedure: $\theta_\mathrm{Hanson}$, $T_\mathrm{Hanson}$}\label{sec:Hanson}
For a single slab, given the outgoing energy, material and thickness, we can find the rms angle in the Gaussian approximation by using the Moli\`ere/Fano/Hanson procedure \cite{mcsbg} with
\begin{equation}\label{eqn:thetaHanson}
\theta_\mathrm{Hanson}(E,M_1,x)\;=\;\chi_c\sqrt{B-1.2}/\sqrt{2}
\end{equation}
{\em We use this angle as our standard of comparison for all rms angles obtained by integrating scattering powers over single slabs.} It is Kanematsu's $\theta_\mathrm{MH}$ \cite{kanematsu08}. We have given it the longer name to emphasize that it is not derived from a scattering power. Indeed, we now derive a scattering power from it by numerical differentiation\footnote{~We use the single sided derivative because {\em adding} even a small increment to $x$ gets us into trouble for near stopping length slabs, where $\theta_\mathrm{Hanson}$ levels off \cite{mcsbg} and $T_\mathrm{Hanson}=0$.}
\begin{equation}\label{eqn:THanson}
\frac{d\theta_\mathrm{Hanson}^2}{dx}\;\equiv\;T_\mathrm{Hanson}(E,M_1,x)\;=\;\lim_{\Delta x\rightarrow0}\;\frac{\theta_\mathrm{Hanson}^2(E,M_1,x)-\theta_\mathrm{Hanson}^2(E,M_1,x-\Delta x)}{\Delta x}
\end{equation}
Though we call this $T_\mathrm{Hanson}$ it is not a scattering power in the usual sense. Evaluating it requires a lengthy procedure rather than a simple formula. However it is the `correct' $T$ for single slabs and we will use it later to derive an improved scattering power.\footnote{~Eq.\;(\ref{eqn:THanson}) may be compared with  Eq.\;(11) of Li and Rogers \cite{li} which, it seems to us, defines an average rather than an instantaneous rate of change.}

\subsection{The Single Scattering Correction}\label{sec:nonLocal}
Rather than thinking of a proton {\em entering} a given slab at some energy, let us focus on the proton at a depth $x$ or equivalently, a proton {\em leaving} a single slab of thickness $x$. Assume the material is Be and the energy at $x$ is 20\;MeV. A `local' formula for $T$ uses only those two facts. But there is another parameter: the amount of material overlying $x$. Our 20\;Mev might for instance represent an incident 23.7\,MeV proton passing through an overlying 1\,mm of Be ($x/R_1=0.264$), or a 102\;MeV proton passing through an overlying 5\;cm ($x/R_1=0.95$). Obviously the mean squared MCS angle itself depends not only on $E(x)$ and $M(x)$ but on the quantity of `MCS buildup' material $x/R_1$ traversed to get to $x$. Does $T$, the rate of change of mean squared angle with $x$, also depend on $x/R_1$?

\begin{figure}[p]
\centering\includegraphics[width=6in,height=4.66in]{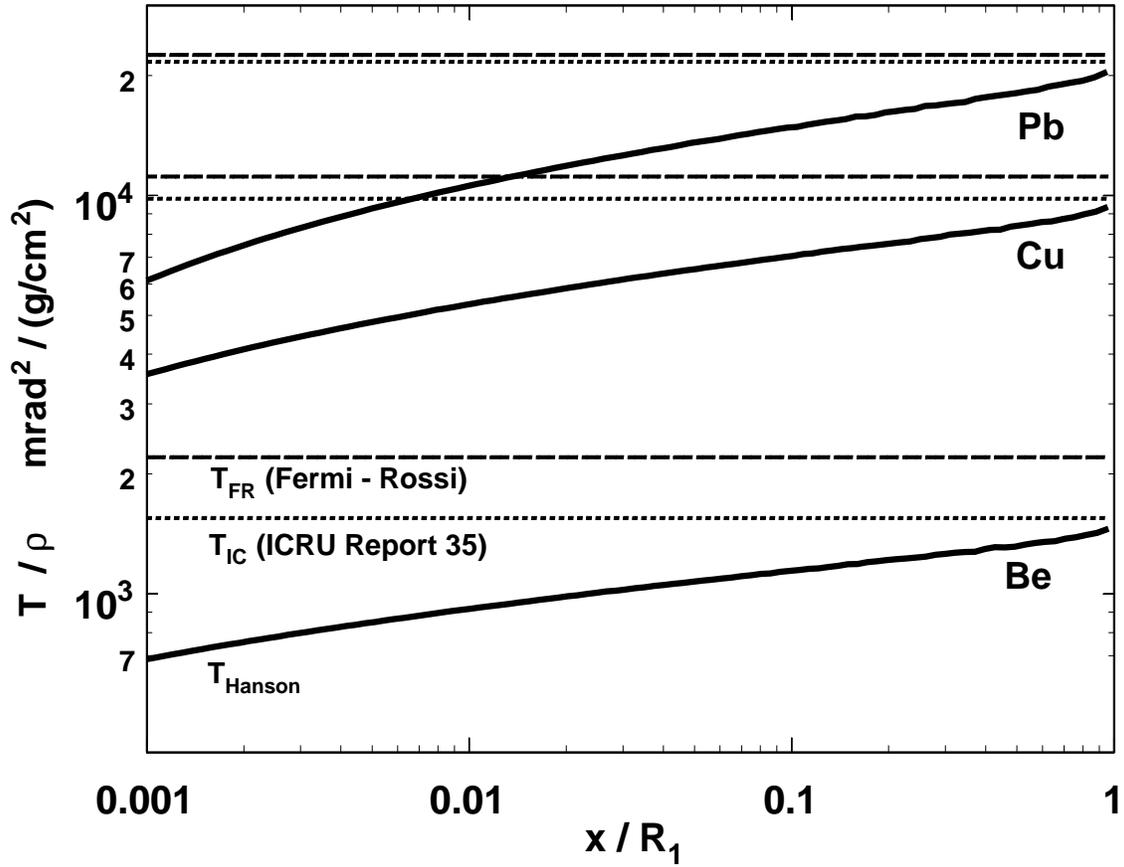} 
\caption{Projected mass scattering powers $T/\rho$ at 20\,Mev in Be, Cu and Pb vs. normalized overlying slab thickness $x/R_1$. Formulas for $T_\mathrm{FR}$, $T_\mathrm{IC}$ and $T_\mathrm{Hanson}$ are given in the text.\label{fig:nonLocal}}
\end{figure}

It does. Figure\;\ref{fig:nonLocal} shows mass scattering power $T/\rho$, computed three different ways for three materials at 20\;MeV, as a function of normalized overlying material $x/R_1$.\footnote{~See Appendix\;\ref{sec:computational} for details on how we computed examples, figures and tables.} $T_\mathrm{Hanson}$ (\ref{eqn:THanson}) is the `correct' scattering power. It is obviously nonlocal: it depends on $x/R_1$. $T_\mathrm{FR}$ and $T_\mathrm{IC}$ (discussed later) are local: they do not depend on $x/R_1$. Of the two, $T_\mathrm{IC}$ has the more accurate material dependence suggesting that it, with some logarithmic function of $x/R_1$ for the single scattering correction, might lead to an accurate approximation to $T_\mathrm{Hanson}$.

\subsection{Highland's Formula: $\theta_\mathrm{Highland}$}\label{sec:Highland}
Highland \cite{highland}, in order to simplify MCS calculations for the high-energy physicist, parameterized Moli\`ere/Bethe/Hanson theory \cite{mcsbg} and obtained the elegantly simple formula\footnote{~The `$E_B$ constant', strictly following Highland's paper, should be $17.5\times1.125/\sqrt{2}=13.92$. However, the 1986 Particle Properties Data Book gave 14.1 which, for whatever reason, we \cite{mcsbg} used in our comparison with experimental data. That has since become the accepted value, used by both Schneider et al. \cite{schneider01} and Kanematsu \cite{kanematsu08}.}
\begin{equation}\label{eqn:Highland}
\theta_0\;=\;\frac{14.1\,\mathrm{MeV}}{p_1v_1}\sqrt{\frac{x}{X_0}}
  \left(1+\frac{1}{9}\log_{10}\frac{x}{X_0}\right)
\end{equation}
in which the single scattering correction, in parentheses, is evident. He assumed a slab that is finite but sufficiently thin that $pv$ does not decrease much ($x\ll R_1$). Thus (\ref{eqn:Highland}) is already an integral expression. It cannot be applied to a thicker slab by interpreting $x$ as a step size $\Delta x$, using (\ref{eqn:Highland}) for each step, and adding in quadrature: the sum decreases indefinitely with the number of steps. 

\begin{figure}[p]
\centering\includegraphics[width=4.72in,height=3.5in]{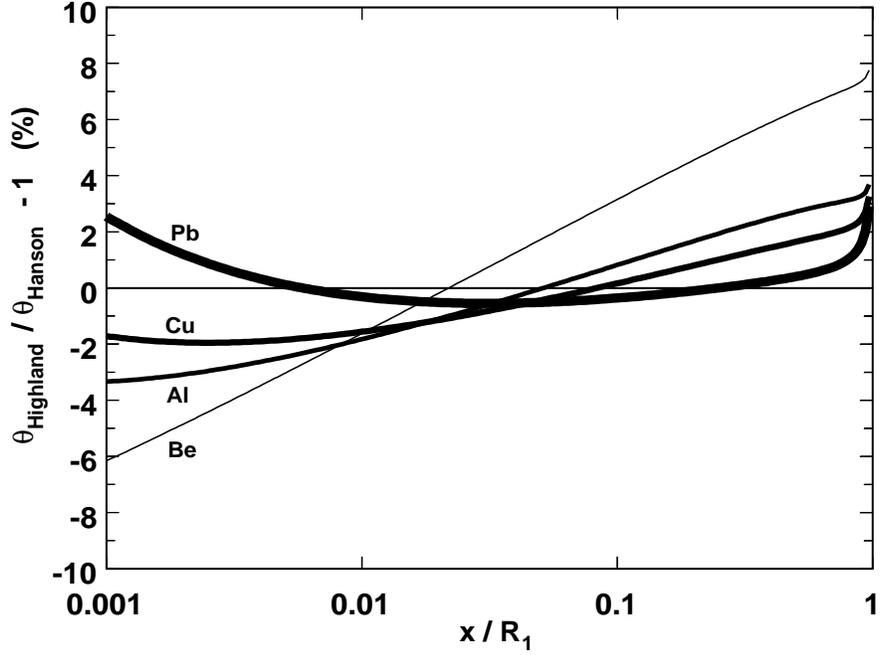} 
\caption{Deviation from $\theta_\mathrm{Hanson}$ of $\theta_\mathrm{Highland}$ computed from the generalized Highland formula (\ref{eqn:GenHighland}), for four scattering materials. The incident proton energy is 158.6\,MeV.
\label{fig:ThH}}
\end{figure}

We circumvented that problem in \cite{mcsbg} by arbitrarily removing the logarithmic term from the sum (or integral) and proposing a generalized Highland formula
\begin{equation}\label{eqn:GenHighland}
\theta_\mathrm{Highland}\;=\;\left(1+\frac{1}{9}\log_{10}\frac{x}{X_0}\right)
\left(\int_0^{x}\left(\frac{14.1\,\mathrm{MeV}}{pv(x')}\right)^2\frac{1}{X_0}\;dx'\right)^{1/2}
\end{equation} 
which reduces to (\ref{eqn:Highland}) for thinner slabs and was found \cite{mcsbg} to agree with measurement almost as well as Moli\`ere/Fano theory. It is Kanematsu's $\theta_{iH}$, which we have given a longer name to emphasize that it is not computed from a scattering power. Figure\,\ref{fig:ThH} compares $\theta_\mathrm{Highland}$ (\ref{eqn:GenHighland}) with $\theta_\mathrm{Hanson}$ (\ref{eqn:thetaHanson}). The behavior for Be reflects the difference between the Moli\`ere/Bethe/Hanson and Moli\`ere/Fano/Hanson forms of the theory. Otherwise, $\theta_\mathrm{Highland}$ is better than the $\pm5$\,\% advertised \cite{highland}, and much easier to compute than the full theory.

\section{Prior Formulas for $T$}\label{sec:review}

At this writing, three local and two nonlocal formulas for $T$ can be found in the literature. We review the first two at some length because Rossi's excellent book \cite{rossiBook} is no longer easily available and because the second, $T_\mathrm{IC}$, is the basis for our improved $T$.

\subsection{Fermi-Rossi: $T_\mathrm{FR}$, $\theta_\mathrm{FR}$\label{sec:TFR}}
Following Rossi \cite{rossiBook} the single scattering probability for a singly charged particle in small angle approximation is \footnote{~We use Moli\`ere's notation $\chi$ for single scattering space angle and related parameters and, in order to follow Rossi more closely, briefly use $\Theta$ for the multiple scattering space angle to distinguish it from $\theta$, the projected angle.}
\begin{equation}\Xi(\chi)\,d\Omega\,dx\;=\;4\,Nr_e^2\,\frac{\rho Z^2}{A}\;\Bigl(\frac{m_e\,c^2}{pv}\Bigr)^2\;
  \frac{1}{\chi^4}\,d\Omega\,dx\label{eqn:Xi}
\end{equation} 
$r_e$ is the classical electron radius, $N$ is Avogadro's number, $A$, $Z$ and $\rho$ are the atomic weight, atomic number and density of the target material and $m_ec^2$ is the electron rest energy. 

In the derivation of (\ref{eqn:Xi}), the target nucleus is modeled as an unscreened point charge. More realistically, the scattering law must depart from $1/\chi^4$ at small angles (distant collisions) of order
\begin{equation}\label{eqn:chi1}
\chi_1\;=\;1.13\alpha Z^{1/3}\left(\frac{m_ec^2}{pc}\right)
\end{equation}
due to screening of the nuclear charge by atomic electrons, and at large angles (close collisions) of order \footnote{~We have kept two changes from ICRU Report~35. The 1.13 comes from the Thomas-Fermi radius of the atom, $r_a=0.885\,a_0\,Z^{-1/3}$ ($a_0=$\;Bohr radius), where Rossi used 1 instead of 0.885. In Eq.(\ref{eqn:chi2}) ICRU35 rounded Rossi's (1/0.49) to 2.}
\begin{equation}\label{eqn:chi2}
\chi_2\;=\;\frac{2}{\alpha A^{1/3}}\left(\frac{m_ec^2}{pc}\right)
\end{equation}
due to the finite size of the nucleus. $\alpha$ is the fine structure constant.

Rossi now assumes that the value of $\overline{\Theta^2}$ at $x+dx$ equals its value at $x$ plus the mean squared space angle of scattering in $dx$. This step is equivalent to assuming the process is {\em exactly} Gaussian and ignoring the single scattering correction. It leads to
\begin{equation}
d\overline{\Theta^2}\;=\;\rho dx\;\int_0^{2\pi}\int_{\chi_1}^{\chi_2}\chi^2\,\Xi(\chi)\;d\Omega_{\chi}
\end{equation}
Rossi now defines 
\begin{equation}\label{eqn:thetassq}
\Theta_s^2\;\equiv\;\frac{1}{\rho}\frac{d\overline{\Theta^2}}{dx}\;=
  \;\int_0^{2\pi}\int_{\chi_1}^{\chi_2}\chi^2\Xi(\chi)\;d\Omega_{\chi}
\end{equation}
Later, Brahme \cite{brahme} used $(T/\rho)$ and the term `mass scattering power' for $\Theta_s^2$. We remind the reader that this and other early uses of $(T/\rho)$ refer to the rate of change of {\em space} angle unlike later treatments and the present paper.

To perform the integral in (\ref{eqn:thetassq}) analytically we must assume some simple behavior of $\Xi(\chi)$ below $\chi_1$ and above $\chi_2$. Rossi does this two different ways. In the first, Rossi assumes $\Xi$ is zero in both regions. After integrating, simplifying and introducing\footnote{~Here we switch back to projected angle $\theta$.} 
\begin{equation}
E_s\;\equiv\;\left(\frac{2\pi}{\alpha}\right)^{1/2}m_ec^2\;=\;15.0\mathrm{\;MeV}
\end{equation}
he obtains
\begin{equation}\label{eqn:TR}
T_\mathrm{FR}\;=\;\left(\frac{E_s}{pv}\right)^2\;\frac{1}{X_0}
\end{equation}
$X_0$ (cm) is the {\em radiation length} of the material. For a sufficiently thin single slab $pv$ does not change much, the integral \[A_0\;=\;<\theta^2>\;=\;\int_0^xT(x')dx'\]
is trivial and
\begin{equation}\label{eqn:thetaFR}
\theta_\mathrm{FR}\;=\;\frac{E_s}{p_1v_1}\;\sqrt{\frac{x}{X_0}}
\end{equation}
the well known Rossi formula. For thicker slabs, $pv$ can be related to $x'$ as long as we know the range-energy relation in the material, and the integral is performed numerically.\footnote{~See Appendix \ref{sec:computational} for details.} Figure \ref{fig:ThR} compares $\theta_\mathrm{FR}$, so computed, to $\theta_\mathrm{Hanson}$. $\theta_\mathrm{FR}$ is far too large for thin scatterers, by an amount which depends on material.

\subsection{ICRU Report 35: $T_\mathrm{IC}$, $\theta_\mathrm{IC}$}\label{sec:TIC}
Rossi next cites a much better approximation, assuming that $\Xi(\chi)$ behaves as $1/(\chi^2+\chi_1^2)^2$ at small angles (leveling off at small $\chi$ rather than suddenly vanishing). Eq.\,(\ref{eqn:thetassq}) then gives
\begin{equation}\label{eqn:TRossi2}
T_\mathrm{IC}\;=\;\alpha N r_e^2\left(\frac{E_\mathrm{s}}{pv}\right)^2\frac{\rho Z^2}{A}\left\{\log\left(1+\left(\frac{\chi_2}{\chi_1}\right)^2\right)-1+\left(1+\left(\frac{\chi_2}{\chi_1}\right)^2\right)^{-1}\right\}
\end{equation}
This is in essence the scattering power given in ICRU Report\;35 \cite{icru35}, except that the version given there takes advantage of a chance cancelation with $m_ec^2$ and therefore, unlike (\ref{eqn:TRossi2}), {\em only applies to electrons}. Nevertheless, we shall call it $T_\mathrm{IC}$ to distinguish it from $T_\mathrm{FR}$.

For protons, $T_\mathrm{IC}$ can be simplified considerably. When $\chi_2$ (\ref{eqn:chi2}) comes out larger than 1\;radian it should be truncated to 1 \cite{rossiBook}. For radiotherapy protons that never happens. The worst case is 3\;Mev protons in Be where $\chi_2=0.913$\;rad. We can therefore simplify $\chi_2/\chi_1$, canceling the $p$ dependence. If we also ignore the rightmost term in $\chi_2/\chi_1$, which is always much less than 1, and introduce a scattering length $X_S$ defined by
\begin{equation}\label{eqn:XS}
\frac{1}{\rho X_S}\;\equiv\;\alpha N r_e^2\,\frac{Z^2}{A}\left\{2\log(33219\,(AZ)^{-1/3})-1\right\}
\end{equation}
we find for radiotherapy protons, 3 to 300\,MeV,
\begin{equation}\label{eqn:TI}
T_\mathrm{IC}\;=\;
\left(\frac{E_s}{pv}\right)^2\;\frac{1}{X_S}
\end{equation}
identical in form to $T_\mathrm{FR}$ (\ref{eqn:TR})

For compounds or mixtures, any scattering power obeys a Bragg rule. Atoms act independently, and the compound or mixture is equivalent to very thin sheets of each constituent in the correct proportion.\footnote{~On fundamental grounds one would expect the Bragg rule for scattering powers to be very much better than for stopping powers, where (through $I$) there is some sensitivity to molecular binding \cite{icru49}.} That picture leads to
\begin{equation}\label{eqn:XSBragg}
\frac{1}{\rho X_S}\;=\;\sum_iw_i\left(\frac{1}{\rho X_S}\right)_i
\end{equation}
where $w_i$ is the fraction by weight of the i$^\mathrm{th}$ constituent. $X_0$ obeys a similar formula. Table \ref{tbl:LS} compares $X_S$ with $X_0$ for a few materials. Figure \ref{fig:ThI} compares $\theta_\mathrm{IC}$, obtained by 
\begin{table}[h]
\begin{center}\begin{tabular}{lrrrrrr}\vspace{3pt}
&\multicolumn{1}{c}{Be}&Lexan&\multicolumn{1}{c}{H$_2$O}&\multicolumn{1}{c}{Al}&
\multicolumn{1}{c}{Cu}&\multicolumn{1}{c}{Pb}\\
$\rho X_S$\quad(g/cm$^2$)&92.60&55.05&46.88&28.75&14.62&6.62\\
$\rho X_0$\quad(g/cm$^2$)&65.19&41.46&36.08&24.01&12.86&6.37\\
$X_S/X_0$&1.420&1.328&1.299&1.197&1.137&1.040\\
\end{tabular}\end{center}
\caption{Comparison of scattering length $X_S$ with radiation length $X_0$ for six materials.\label{tbl:LS}}
\end{table}
integrating $T_\mathrm{IC}$, with $\theta_\mathrm{Hanson}$. Material dependence is greatly improved over $T_\mathrm{FR}$, but the large error for thin scatterers remains.

\begin{figure}[p]
\centering\includegraphics[width=4.54in,height=3.5in]{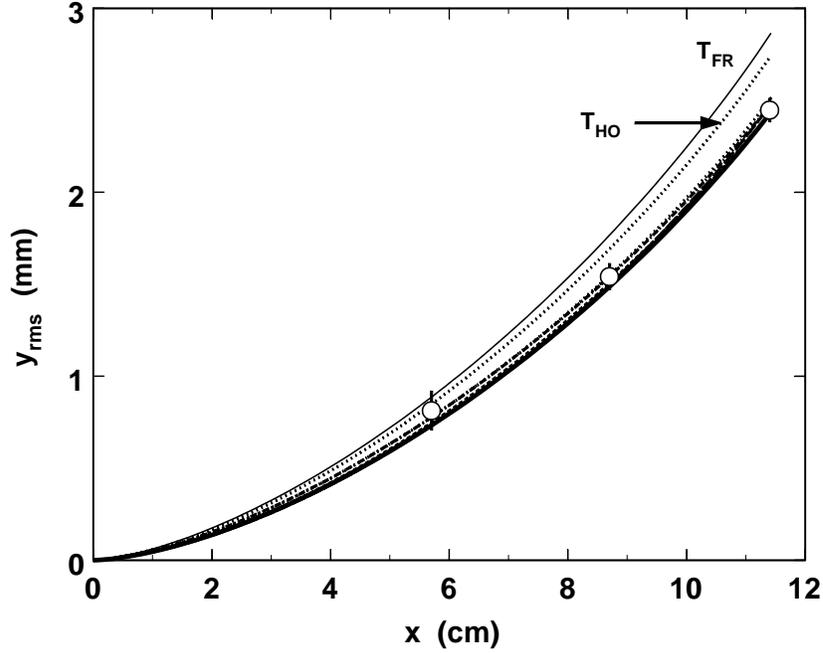} 
\caption{Spreading of a 127\;MeV proton pencil beam in water with experimental data from Preston and Koehler \cite{preston}. $T_\mathrm{HO}$ is taken from Table\;2 of Hollmark et al. \cite{hollmark}. Apart from $T_\mathrm{FR}$ and $T_\mathrm{HO}$, calculations based on the other scattering powers are barely distinguishable.\label{fig:YrmsWATER}}
\end{figure}

Hollmark et al. \cite{hollmark} use $T_\mathrm{IC}$ for protons and heavy ions. However, the formula they quote (their Eq.\;(24)) is only valid for electrons, and they introduce an effective charge factor $Z_\mathrm{P,eff}$ without comment.\footnote{~If we use the Barkas formula $Z_\mathrm{eff}=Z(1-\exp(-125\beta Z^{-2/3}))$ \cite{kraft,barkas} then $Z_\mathrm{P,\,eff}=1$ for radiotherapy protons.} Their values of $T/\rho$ for protons up to 200\;MeV in water (their Table\;2) are larger than ours by a factor 1.19, perhaps due to incorrect application of the Bragg rule. Our Figure\;\ref{fig:YrmsWATER} seems to confirm that discrepancy. A footnote to a more recent paper \cite{hollmark08} by the same group corrects the headings of Table\;2 but does not mention any numerical error. 

\subsection{Linear Displacement: $T_\mathrm{LD}$, $\theta_\mathrm{LD}$}\label{sec:TLD}
\begin{figure}[p]
\centering\includegraphics[width=4.72in,height=3.5in]{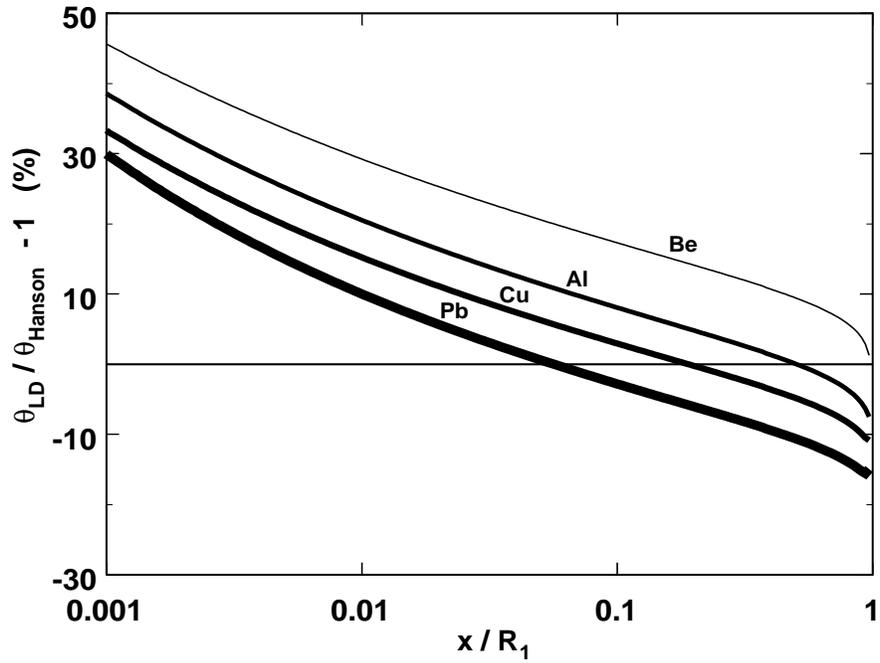} 
\caption{Deviation from $\theta_\mathrm{Hanson}$ of $\theta_{LD}$ computed from $T_\mathrm{LD}$, Kanematsu's `linear displacement' scattering power \cite{kanematsuLD}, for four scattering materials. The incident proton energy is 158.6\,MeV.
\label{fig:ThLD}}
\end{figure}

In a recent note \cite{kanematsuLD} Kanematsu proposes a simple scattering power for protons and heavy ions in tissue-like matter. He uses water as a reference material. For protons, his formula reduces to
\begin{equation}\label{eqn:TLD}
T_\mathrm{LD}\;=\;1.00\times10^{-3}\;\frac{\rho_x}{R_W}
\end{equation}
In a single slab of tissue-like material $M$
\begin{eqnarray}
\rho_x&\equiv&X_\mathrm{0W}/X_\mathrm{0M}\\
R_W&\equiv&R_\mathrm{1W}-\rho_s\;x\\
\rho_s&\equiv&S_\mathrm{M}/S_\mathrm{W}\label{eqn:Sratio}
\end{eqnarray}
where $W$ stands for water, $S$ is stopping power (MeV/cm) and $R_W$ (cm) is the proton's residual range in water at depth $x$. In the interest of a more complete survey we will, in what follows, explore the behavior of $T_\mathrm{LD}$ for non tissue-like materials, even though it may not work well for those. We call it the `Linear Displacement' scattering power after Kanematsu's derivation. For a single slab it can be integrated analytically giving
\begin{equation}
A_0\;=\;<\theta_\mathrm{LD}^2>\;=\;\int_0^xT_\mathrm{LD}(x')dx'\;=\;1.00\times10^{-3}\;\frac{\rho_x}{\rho_s}\;\ln\left(\frac{R_\mathrm{1W}}{R_W}\right)
\end{equation}
A plot of $\theta_\mathrm{LD}$ vs. $x/R_1$ (Figure \ref{fig:ThLD}) is reminiscent of $\theta_\mathrm{FR}$ except near end-of-range.\footnote{~For the sake of uniformity, Figure \ref{fig:ThLD} was obtained by integrating (\ref{eqn:TLD}) numerically. A small difficulty arises for non tissue-like material because the stopping power ratio (\ref{eqn:Sratio}) is then somewhat energy dependent. That introduces some irregularity near end-of-range, which we minimized by evaluating $\rho_s$ at $0.64\times E_1$.} Indeed $T_\mathrm{LD}$ is a variant of $T_\mathrm{FR}$ with three changes: the weak {\O}ver{\aa}s approximation (\ref{eqn:overasWeak}) is used for $pv$, $E_s$ is adjusted downward, and water is used as a reference material. To show this consider a single slab of water. Then
\begin{eqnarray}
T_\mathrm{FR,W}&=&\frac{E_s^2}{X_\mathrm{0W}}\;\frac{1}{(pv)^2}\;\approx\;\frac{E_s^2}{X_\mathrm{0W}}\;\frac{R_\mathrm{1W}}{(pv)_1^2}\;\frac{1}{R_\mathrm{1W}-x}\hbox{\qquad from (\ref{eqn:overasWeak})}\nonumber\\
&=&\frac{15^2}{36.08}\times(2\times10^{-4})\;\frac{1}{R_\mathrm{W}}\;=\;1.25\times10^{-3}\;\frac{1}{R_\mathrm{W}}\label{eqn:result}
\end{eqnarray}
if we evaluate $R_\mathrm{1W}/(p_1v_1)^2$ at 158.6\,MeV (it is insensitive to $E_1$). When  1.25 is reduced to 1.00 and (\ref{eqn:result}) is generalized to other materials by introducing ratios to water, (\ref{eqn:TLD}) follows.

Kanematsu \cite{kanematsuLD} freely admits that $T_\mathrm{LD}$ is semi-empirical, but his derivation \cite{kanematsu08,kanematsuLD} is complicated and it is somewhat unclear exactly where the downward adjustment of $E_s$ takes place. It is probably mostly present in his Eq.\;(6) \cite{kanematsuLD} which flows from Highland's equation which in turn is an empirical fit to Moli\`ere/Bethe/Hanson theory \cite{highland}.

There is precedent for adjusting $E_s$ downward to improve the performance of $T_\mathrm{FR}$ over a limited range of normalized thickness. Soukop et al. multiplied $E_s$ by a factor 0.8, said \cite{soukop} to have been obtained from a fit to Geant\,4, in their `corrected Rossi' formula. Kanematsu's reduction factor is only $(1/1.25)^{1/2}=0.89$. 

\subsection{{\O}ver{\aa}s-Schneider: $T_\mathrm{{\O}S}$, $\theta_\mathrm{{\O}S}$\label{sec:TOS}}
Schneider et al. \cite{schneider01} describe a scattering power based on $T_\mathrm{FR}$ with a nonlocal correction factor in the form of an analytical function of a new normalized variable $t$. For a single slab, $t(x)$ is just $x/R_1$. For mixed slabs 
\begin{equation}\label{eqn:t} 
t(x)\equiv x/R(E_1,M(x))
\end{equation} 
where $M(x)$ is the current material. $t$ is a discontinuous function of $x$. It is that normalized depth which would be obtained if the protons were degraded from $E_1$ to $E(x)$ in the {\em current} material.

\begin{figure}[p]
\centering\includegraphics[width=4.72in,height=3.5in]{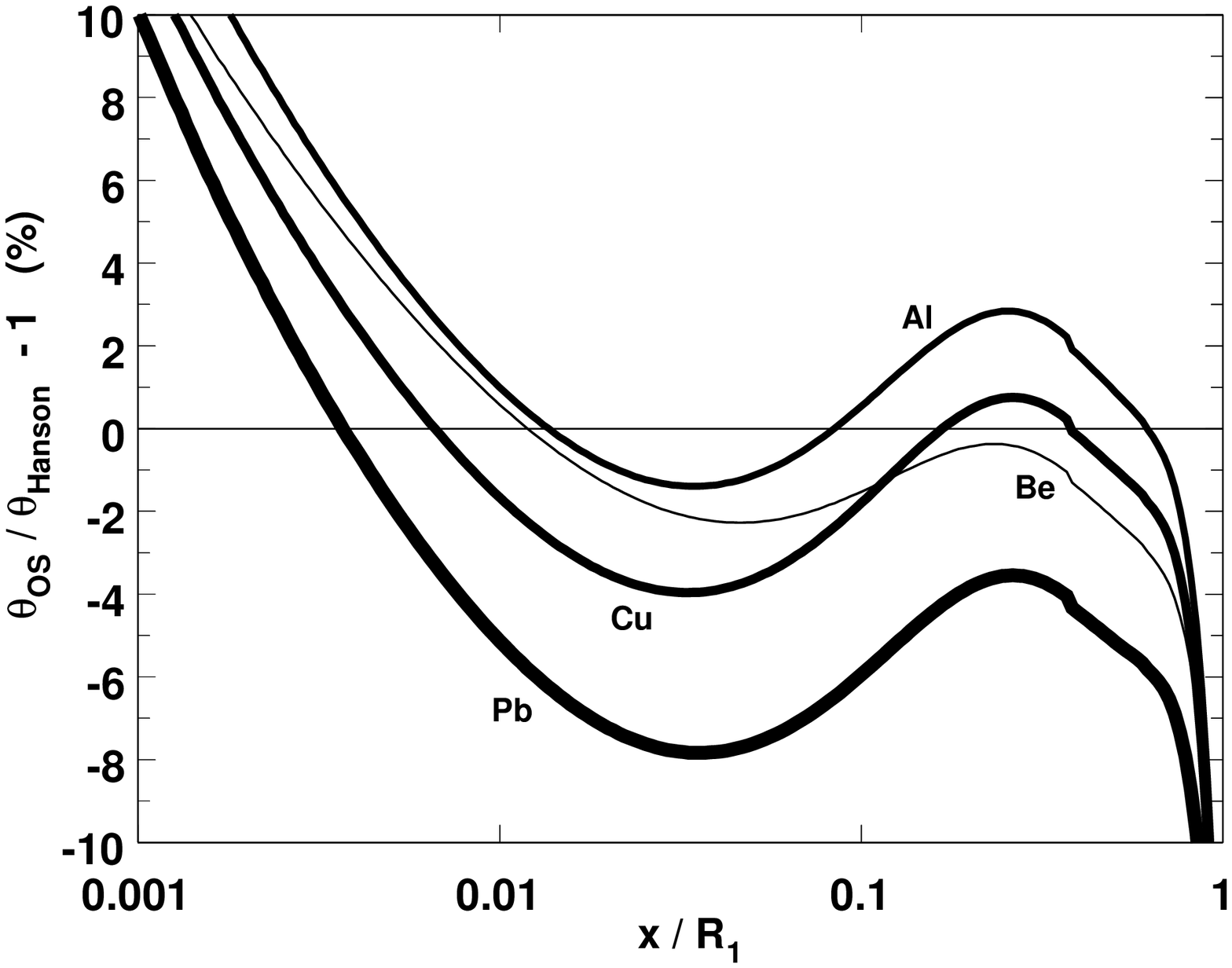} 
\caption{Deviation from $\theta_\mathrm{Hanson}$ of $\theta_\mathrm{{\O}S}$ computed from $T_\mathrm{{\O}S}$, the scattering power of Schneider et al. \cite{schneider01}, for four scattering materials. The incident proton energy is 158.6\,MeV.
\label{fig:ThS}}
\end{figure}

They derive their function of $t$ by fitting a large body of single slab experimental data for $\theta_{rms}$ at 158.6\;Mev \cite{mcsbg} with a two parameter analytic function of $t$ (their Eq.\,(10)). Differentiating, they find\footnote{~We have corrected two typographic errors.}
\begin{eqnarray}\label{eqn:TSC}
\lefteqn{T_{{\O}S}\;=\;\left(\frac{19.9\mathrm{\,MeV}}{p_1v_1}\right)^2\;\frac{1}{X_0}\times}\nonumber\\
&&(1-t)^{-(1+k)}\left\{c_0\;+\;c_1\left(t-\frac{1}{2}\right)^4
\;+\;\frac{4c_1}{k}\left(t-\frac{1}{2}\right)^3(1-t)\left(1-(1-t)^k\right)\right\}
\end{eqnarray}
with fitted constants $k(X_0)$ (\ref{eqn:k}) and
\begin{eqnarray}
c_0&=&(201/200)\;-\;(23/5000)\rho X_0\label{eqn:c0}\\
c_1&=&-(11/2)\;+\;(43/1000)\rho X_0\label{eqn:c1}
\end{eqnarray}
For mixed slabs, $\rho X_0$ in Eqs.\;(\ref{eqn:k}), (\ref{eqn:TSC}), (\ref{eqn:c0}) and (\ref{eqn:c1}) is the mass radiation length of the {\em current} material. Figure \ref{fig:ThS} shows considerable improvement over the local formulas for $T$ (note change in scale). Oscillations and divergent behavior at the ends are characteristic of polynomial fits. 

Two general remarks. First, Schneider et al. fit experimental data rather than some form of multiple scattering theory. Since theory and measurement agree rather well \cite{mcsbg} this should not have a large effect, but it biases their formula towards the data that happen to be available. Second, their major advance is the introduction of a nonlocal correction based on a generalized definition of normalized depth. The {\O}ver{\aa}s approximation, tightly woven into their formalism to obtain formulas in closed form, is somewhat of a distraction. Similar results could be obtained without it. 

\subsection{Differential Highland: $T_\mathrm{dH}$, $\theta_\mathrm{dH}$}\label{sec:TdH} 
\begin{figure}[p]
\centering\includegraphics[width=4.72in,height=3.5in]{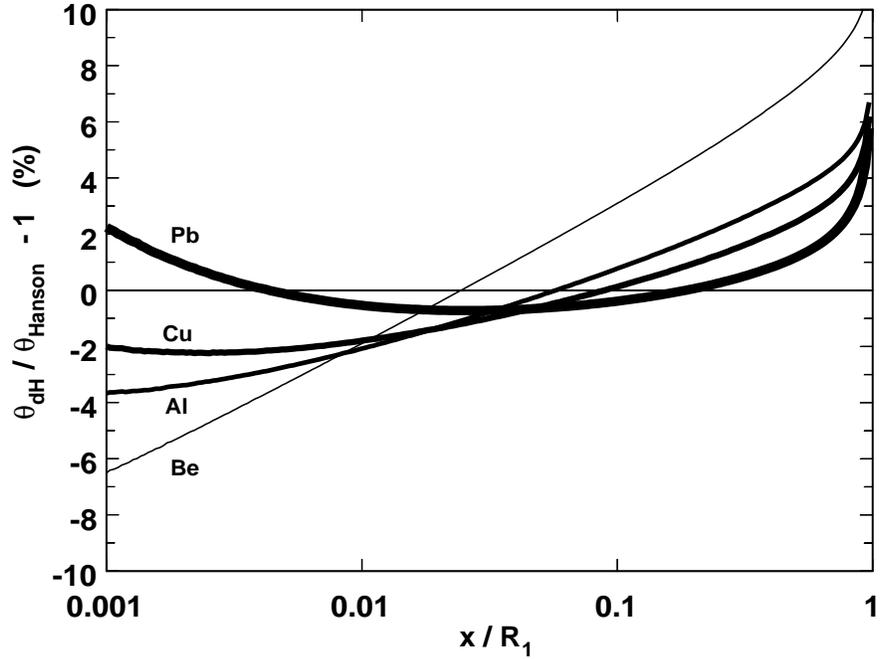} 
\caption{Deviation from $\theta_\mathrm{Hanson}$ of $\theta_\mathrm{dH}$ computed from $T_\mathrm{dH}$, Kanematsu's `differential Highland' scattering power \cite{kanematsu08}, for four scattering materials. The incident proton energy is 158.6\,MeV.
\label{fig:ThK}}
\end{figure}

Kanematsu \cite{kanematsu08} derives a nonlocal scattering power applicable to mixed slabs which by construction gives the same result as (\ref{eqn:Highland}) for a single thin slab. In his case the nonlocality parameter is depth weighted by inverse radiation length: he generalizes $x/X_0$ to a dimensionless {\em radiative path length}
\begin{equation}
\ell(x)\;\equiv\;\int_0^x\frac{dx'}{X_0(x')}
\end{equation}
and writes a new scattering power as $T_\mathrm{FR}$ times a correction factor
\begin{equation}\label{eqn:TK}
T_\mathrm{dH}\;\equiv\;f_\mathrm{dH}(\ell)\;\left(\frac{E_s}{pv}\right)^2\;\frac{1}{X_0}
\end{equation}
He now constructs $f$ so that, for a single thin slab, the integral of $T_\mathrm{dH}$ will equal $\theta_\mathrm{Highland}$. That requires that the average of $f$ equal the squared ratio of the Highland formula (\ref{eqn:Highland}) to the Rossi formula (\ref{eqn:thetaFR})\footnote{~$\lg\equiv\log_{10}$, $\ln\equiv\log_e$.}
\[
\frac{1}{\ell}\int_0^\ell f_\mathrm{dH}(\ell')d\ell'\;=\;\left(1+\frac{\lg\ell}{9}\right)\left(\frac{14.1\hbox{\;MeV}}{E_s}\right)^2
\]
Differentiating, he finds after some algebra
\begin{equation}\label{eqn:fdH}
f_\mathrm{dH}(\ell)\;=\;0.970\left(1+\frac{\ln\ell}{20.7}\right)\left(1+\frac{\ln\ell}{22.7}\right)
\end{equation}
to be used in (\ref{eqn:TK}). Figure\,(\ref{fig:ThK}) compares $\theta_\mathrm{dH}$, obtained by integrating $T_\mathrm{dH}$, with $\theta_\mathrm{Hanson}$. It indeed behaves very like $\theta_\mathrm{Highland}$ except for thick slabs because the derivation of (\ref{eqn:fdH}) is strictly correct only for thin slabs.

Unlike $T_\mathrm{{\O}S}$ the extension of $T_\mathrm{dH}$ to mixed slabs is totally straightforward, requiring only the new path integral $\ell(x)$.

\section{Improved Nonlocal Formula: $T_\mathrm{dM}$, $\theta_\mathrm{dM}$}\label{sec:TdM}
Instead of $T_\mathrm{FR}$ as a basis let us use $T_\mathrm{IC}$ (\ref{eqn:TI}) which, for radiotherapy protons, is as simple and has better material dependence (Figure \ref{fig:ThI}). From Figure\;\ref{fig:nonLocal}, the correction should be logarithmic in total material overlying the POI.

\begin{figure}[p]
\centering\includegraphics[width=4.52in,height=3.5in]{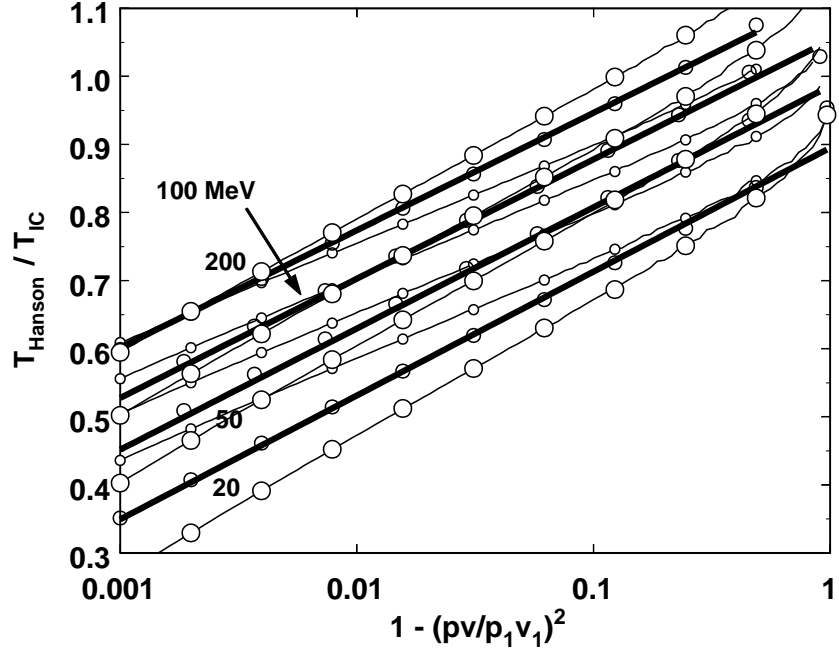} 
\caption{Ratio $T_\mathrm{Hanson}/T_\mathrm{IC}$ as a function of $1-(pv/p_1v_1)^2$ at four proton energies for three materials: Be (small circles), Cu (medium circles) and Pb (large circles). The heavy lines are a bilinear fit to the entire data set.\label{fig:TryPVfit}}
\end{figure}

In the weak {\O}ver{\aa}s approximation (\ref{eqn:overasWeak}) we found that $1-(pv/p_1v_1)^2$, which depends only on local energy $E$ and incident energy $E_1$, is a reasonably good proxy for normalized depth $x/R_1$ for all materials and energies of interest over three orders of magnitude (Figure \ref{fig:PVfuncVsTnLL}). Let us therefore compute and plot the ratio of $T_\mathrm{Hanson}$, the ideal scattering power, to $T_\mathrm{IC}$, our proposed basis, for $0.001\le 1-(pv/p_1v_1)^2\le0.97$ and $E_1\le300$\,Mev, for several values of the energy $E$ at the POI and several materials (our usual Be, Cu and Pb). For each point we compute the exact $x/R_1$ for that material, without relying on the {\O}ver{\aa}s approximation, then use (\ref{eqn:THanson}). The result is shown in Figure\;\ref{fig:TryPVfit}.

An adequate fit to these data, also shown in Figure\;\ref{fig:TryPVfit}, is a linear polynomial in $\lg\,(1-(pv/p_1v_1)^2)$ whose two coefficients are in turn linear in $\lg\,(pv)$. Defining a new `differential Moli\`ere' scattering power and writing everything out we have  
\begin{equation}\label{eqn:TdM}
T_\mathrm{dM}\;=\;f_{dM}(pv,p_1v_1)\times\left(\frac{E_s}{pv}\right)^2\;\frac{1}{X_S}
\end{equation}
where
\begin{equation}\label{eqn:fdM}
f_\mathrm{dM}\;\equiv\;0.5244+0.1975\lg(1-(pv/p_1v_1)^2)+0.2320\lg(pv)-0.0098\lg(pv)\lg(1-(pv/p_1v_1)^2)
\end{equation}
Although the weak {\O}ver{\aa}s approximation suggested the form of (\ref{eqn:fdM}) the final result is an independent fit. It does not depend on the accuracy of either (\ref{eqn:overas}) or (\ref{eqn:overasWeak}), or on the scattering material. 

\begin{figure}[p]
\centering\includegraphics[width=4.72in,height=3.5in]{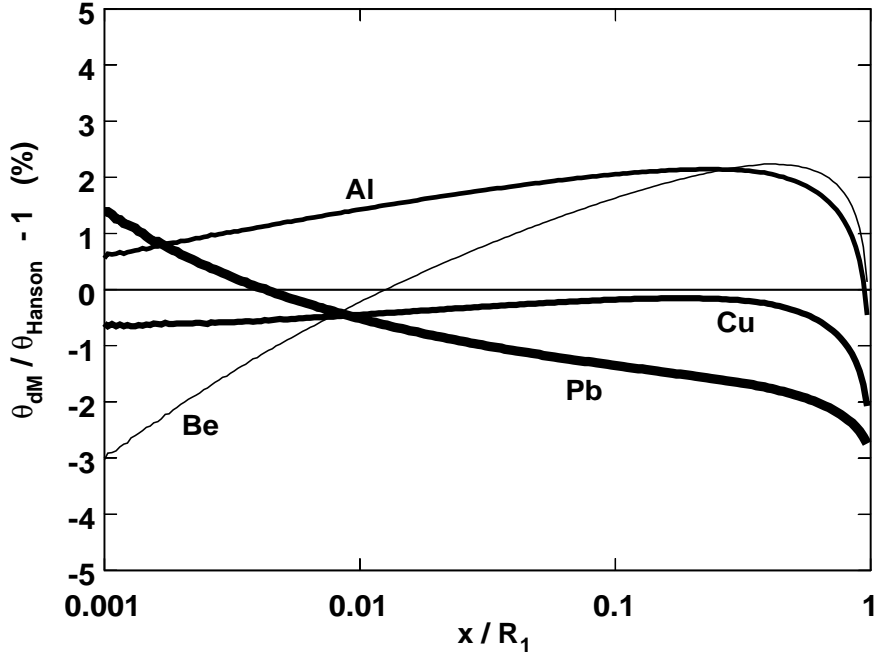} 
\caption{Deviation from $\theta_\mathrm{Hanson}$ of $\theta_\mathrm{dM}$ computed from $T_\mathrm{dM}$, the scattering power proposed in the present work, for four scattering materials. The incident proton energy is 158.6\,MeV.
\label{fig:ThP}}
\end{figure}

Figure\,\ref{fig:ThP} compares $\theta_\mathrm{dM}$ obtained by integrating $T_\mathrm{dM}$ with $\theta_\mathrm{Hanson}$ (note the vertical scale).

The reader will object that the coefficients of (\ref{eqn:fdM}), here given to excessive precision, are arbitrary. We could have chosen different sets of materials or energies. That is perfectly true, but it is also true of $T_\mathrm{{\O}S}$ and $T_\mathrm{dH}$. The former is a fit to a specific data set for specific materials at a specific energy. The latter stems from Highland's formula, also a fit albeit to theory. That said, we have tried other materials and energies without much change in the general appearance of Figure\,\ref{fig:ThP}.

\begin{figure}[p]
\centering\includegraphics[width=4.72in,height=3.5in]{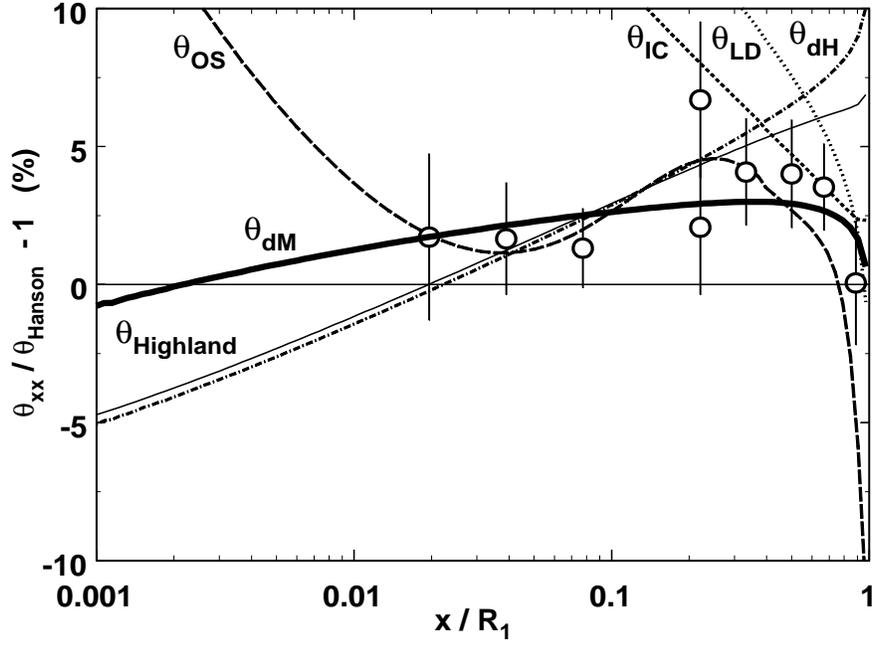} 
\caption{For polystyrene, deviation of $\theta_\mathrm{xx}$ from $\theta_\mathrm{Hanson}$ at 158.6\,Mev incident energy. Data (open circles) taken from \cite{mcsbg}. Each $\theta_\mathrm{xx}$ is the integral of the corresponding $T_\mathrm{xx}$ except $\theta_\mathrm{Highland}$ which is from the generalized Highland formula (\ref{eqn:GenHighland}). $\theta_\mathrm{FR}$ is off scale.\label{fig:ExptFigPoly}}
\end{figure}

\begin{figure}[p]
\centering\includegraphics[width=4.72in,height=3.5in]{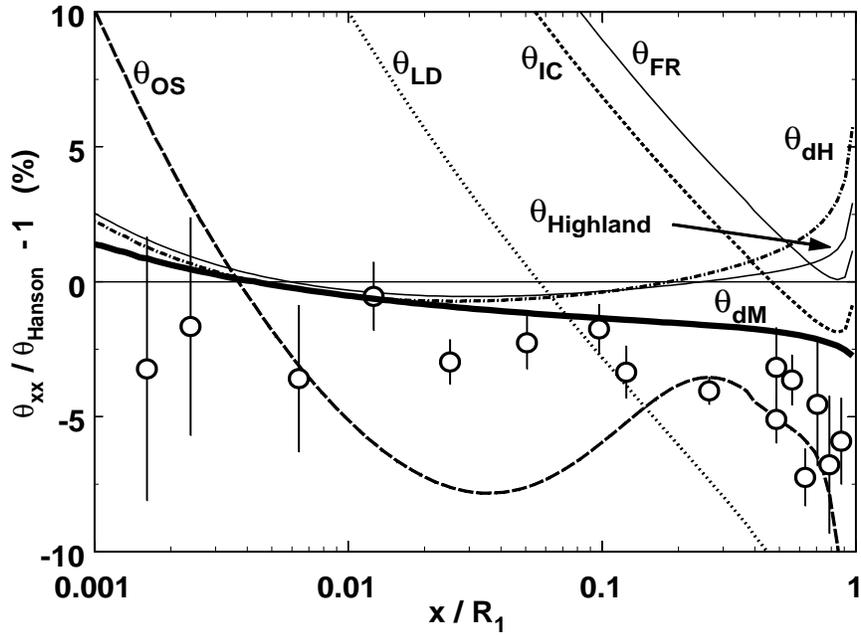} 
\caption{The same as Figure \ref{fig:ExptFigPoly}, for Pb.\label{fig:ExptFigLead}}
\end{figure}

Since $\theta_\mathrm{dM}$ agrees well with  $\theta_\mathrm{Hanson}$ it may be inferred \cite{mcsbg} that it agrees well with experiment for many materials. We will show that directly only for one low-$Z$ and one high-$Z$ material, also using the opportunity for a head-to-head comparison of six scattering powers and the generalized Highland formula. Figure \ref{fig:ExptFigPoly} shows the comparison for polystyrene and  Figure \ref{fig:ExptFigLead} shows it for lead, both with data from \cite{mcsbg}. All three nonlocal $T$'s are better than any local $T$ but $T_\mathrm{dM}$ is the best. It almost agrees with measurement within the experimental error.\footnote{~Reference \cite{mcsbg} already remarked on the fact that experimental data seem a few percent lower than theory for thick Pb slabs. In our present opinion that is more likely due to a systematic experimental error for very large $\theta_{RMS}$ than to a breakdown of Moli\`ere theory, but it is impossible to say for sure.}
  
To facilitate numerical checks we include short tables of $\theta_\mathrm{XX}$ (Table \ref{tbl:theta0Single}) and $(T/\rho)_\mathrm{XX}$ (Table \ref{tbl:ToverRhoSingle}) for single slabs of various materials and normalized thicknesses. We use $T_1=158.6$\,MeV to correspond to \cite{mcsbg} particularly Table\,1. $\theta_\mathrm{Hanson}$ is not given directly there but may be found using (\ref{eqn:thetaHanson}).

The generalization of $T_\mathrm{dM}$ to mixed slabs is the easiest of all. It does not even require an additional path integral. The single scattering correction is a logarithmic function of the fractional decrease in $pv$, with no explicit material dependence, from the incident beam to the point of interest. It diverges, as any single scattering correction must, at $pv/(p_1v_1)=1$ (no overlying material).

\section{Applications}\label{sec:apps}
What kinds of computation are significantly affected by the choice of $T$? If we are only interested in the Gaussian MCS angle itself in a single slab, Figures\;\ref{fig:ThH} through \ref{fig:ThK} and \ref{fig:ThP} through \ref{fig:ExptFigLead} already answer that question. However, in that case we would not use transport theory or $T$ at all but simply evaluate either $\theta_\mathrm{Hanson}$ (\ref{eqn:thetaHanson}) or $\theta_\mathrm{Highland}$ (\ref{eqn:GenHighland}). Let us examine some less trivial cases.

\subsection{Pencil Beam Spreading in a Single Slab}
The archetype for dose reconstruction in the patient is pencil beam spreading in a homogeneous water slab. An early paper by Preston and Koehler \cite{preston} derived a universal formula for beam spreading and compared theory with measurements. Hollmark et al. \cite{hollmark} refer to additional measurements. Figure\;5a of \cite{mcsbg} already suggests that beam spreading in water is insensitive to $T$, comparing two very different models of MCS: Preston and Koehler's which is a local model similar to $T_\mathrm{IC}$, and a beam spreading model based on the nonlocal generalized Highland formula. Insensitivity of beam spreading in water to $T$ implies that almost any $T$ will work reasonably well for dose reconstruction, and is therefore worth a closer look.

\begin{figure}[p]
\centering\includegraphics[width=4.54in,height=3.5in]{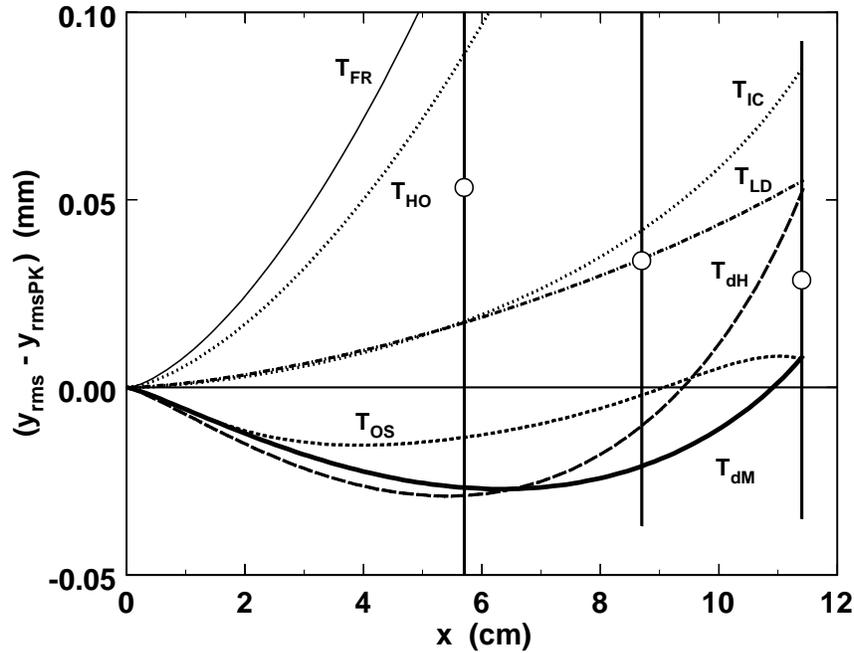} 
\caption{Expanded view of Figure \ref{fig:YrmsWATER} using Preston and Koehler's analytic approximation to beam spreading (see text) as a an arbitrary reference. Line styles are $T_\mathrm{FR}$ light solid, $T_\mathrm{IC}$ dotted, $T_\mathrm{LD}$ dot dash, $T_\mathrm{{\O}S}$ short dash, $T_\mathrm{dH}$ long dash, $T_\mathrm{dM}$ bold solid for this and all following figures.\label{fig:YrmsDifWATER}}
\end{figure}

Figure\;\ref{fig:YrmsWATER} shows spreading of a 127\;MeV pencil beam in water with $y_{rms}=(A_2)^{1/2}$ (Eq.\;\ref{eqn:An}) computed according to all six formulas for $T$ and according to Table\;2 of Hollmark et al. \cite{hollmark} ($T_\mathrm{HO}$). Except $T_{FR}$ and $T_\mathrm{HO}$, all agree with experiment and are barely distinguishable from each other. Figure\;\ref{fig:YrmsDifWATER} is an expanded version where we plot the difference between each calculation and Preston and Koehler's formula (Appendix \ref{sec:PK}). The spread, of order 0.1\,mm, would be negligible in any dose reconstruction problem.

\begin{figure}[p]
\centering\includegraphics[width=4.7in,height=3.5in]{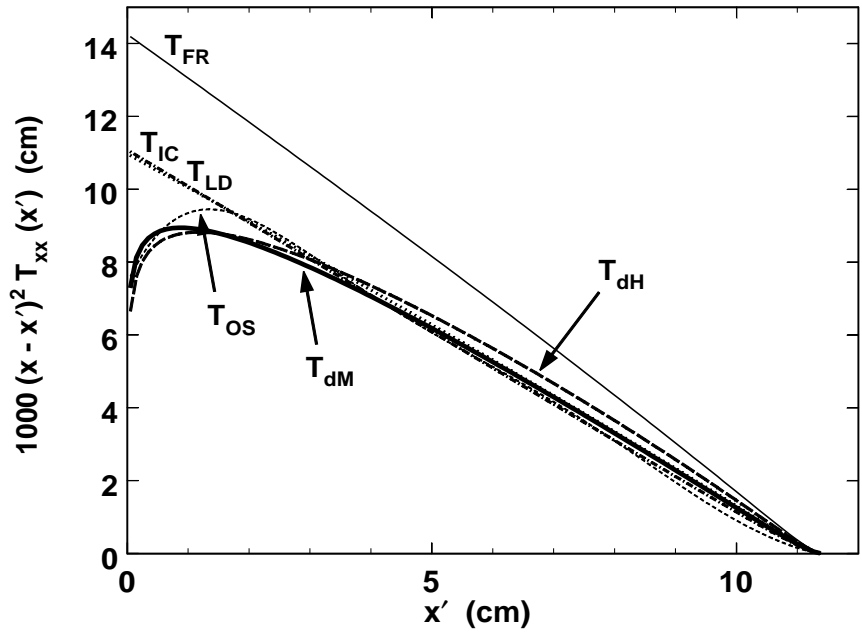} 
\caption{Integrand of $A_2$ at $x=0.97\,R_1$ for 127\;Mev protons incident on water.\label{fig:A2integrandWater097}}
\end{figure}

\begin{figure}[p]
\centering\includegraphics[width=4.7in,height=3.5in]{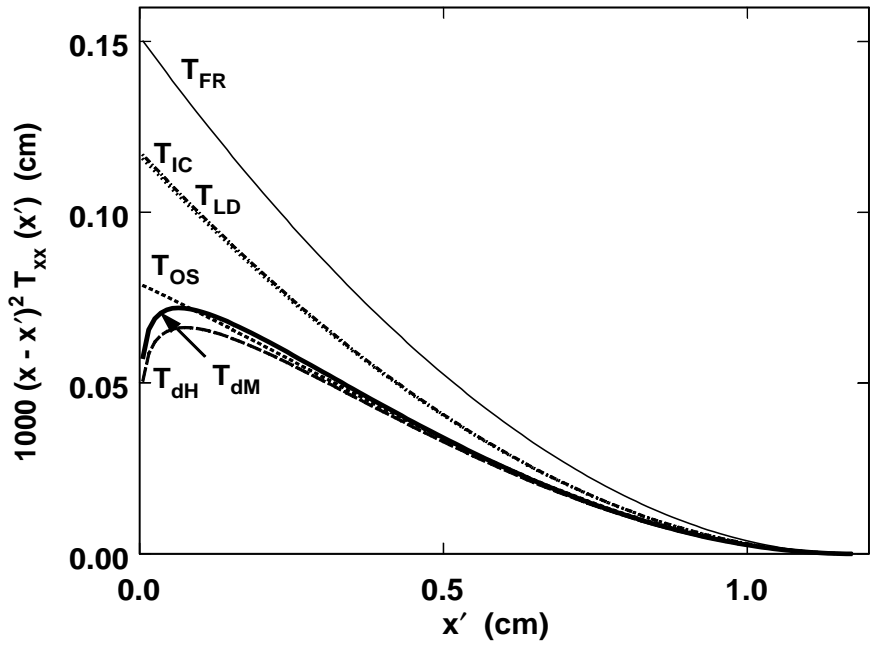} 
\caption{Same as Figure\;\ref{fig:A2integrandWater097} except $x=0.1\,R_1$.\label{fig:A2integrandWater010}}
\end{figure}

To gain some insight into this insensitivity to $T$, Figure\;\ref{fig:A2integrandWater097} shows the integrand of $A_2$ when the POI is at $0.97\,R_1$, near stopping depth. The near linearity of curves for the local formulas ($T_\mathrm{FR}$, $T_\mathrm{IC}$ and $T_\mathrm{LD}$) is a consequence of the {\O}ver{\aa}s approximation. The nonlocal formulas do in fact give a lower result because of the single scattering correction, important for small $x$, but not by much. Except for $T_\mathrm{FR}$, which is too high everywhere, the areas under the curves are nearly the same. That breaks down when we consider the integrand to $x=0.1\,R_1$ (Figure\;\ref{fig:A2integrandWater010}) but by then the {\em absolute} effect is so small as to be negligible.

\begin{figure}[p]
\centering\includegraphics[width=4.5in,height=3.5in]{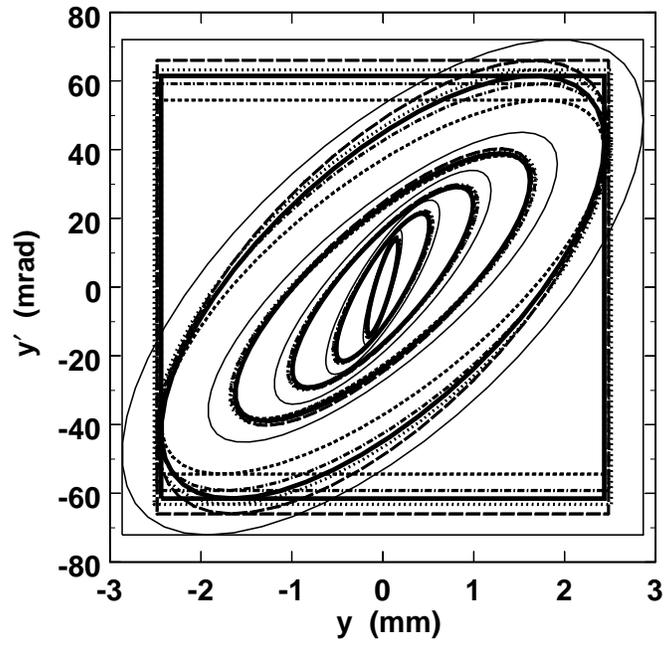} 
\caption{Beam phase space ellipses for 127\;Mev protons incident on water using Fermi-Eyges theory with six scattering powers, at the exit faces of five equal slabs extending to $0.97\;R_1$.\label{fig:ellipsesWater127}}
\end{figure}

\begin{figure}[p]
\centering\includegraphics[width=4.5in,height=3.5in]{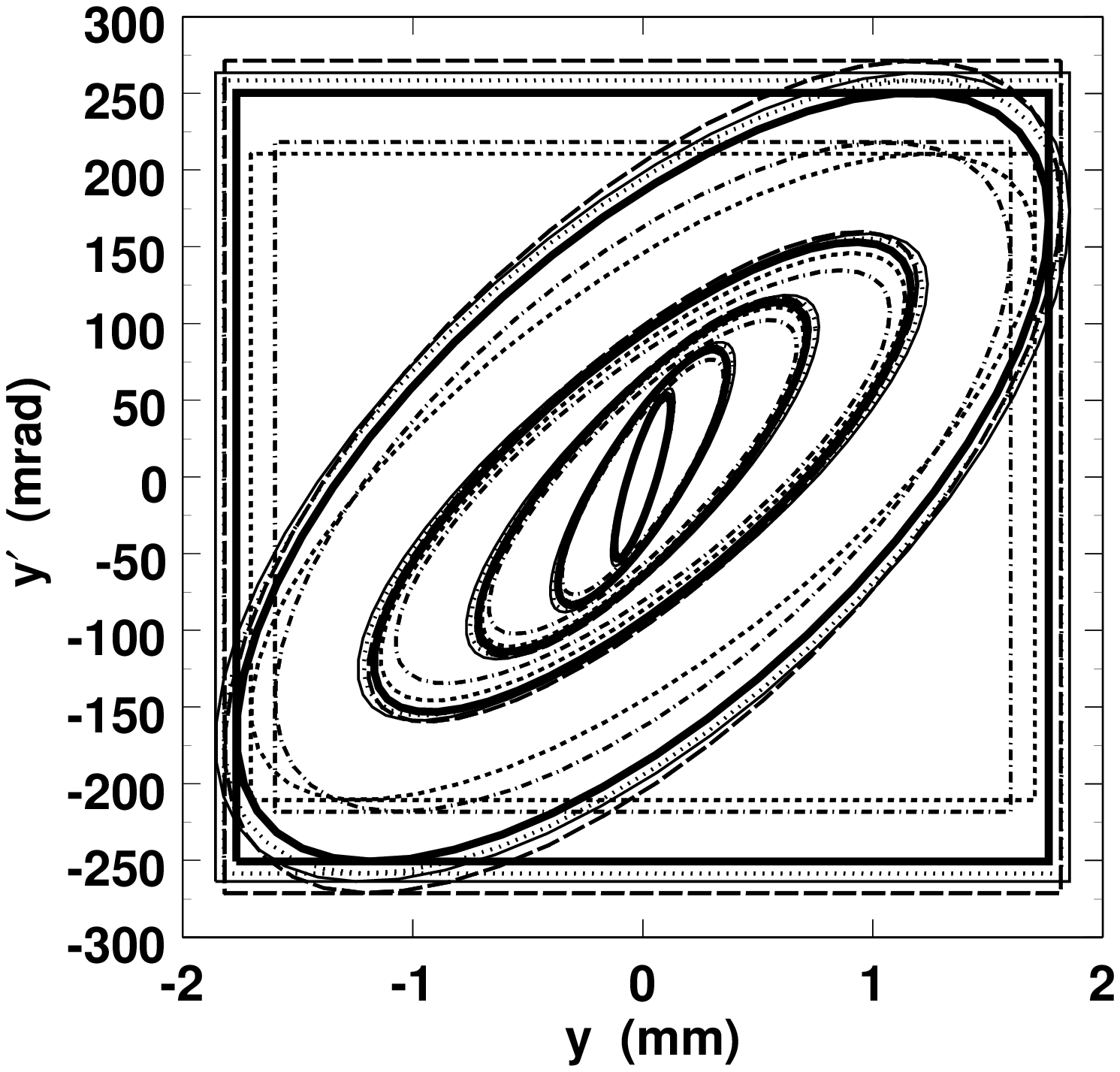} 
\caption{Same as Figure\;\ref{fig:ellipsesWater127} for Pb.\label{fig:ellipsesLead127}}
\end{figure}

It is instructive to look at the same problem in terms of the evolution of the beam phase space ellipse \cite{icru35}. Figure\;\ref{fig:ellipsesWater127} studies the same case as the preceding figures, dividing the water slab up to $x=0.97\,R_1$ into five sub-slabs. We also show bounding boxes for the final slab.\footnote{~Recall that the vertical bound represents $\theta_{rms}$ while the horizontal bound is $y_{rms}$ \cite{icru35}.} The spread of vertical bounds shows that the different $T$'s do give different answers for the final rms angle, but the effect on spatial spreading (horizontal bounds) is almost nil. The same study for Pb, Figure\;\ref{fig:ellipsesLead127}, shows that this is a fortuitous property of tissue-like materials, presumably due to a particular combination of drift and scatter. Beam spreading in near-stopping high-Z slabs is not entirely academic. In computing collimator scatter, we are basically asking how many protons leave the bore of a Cerrobend or other aperture that, except for MCS, would have stopped.

\subsection{MCS Angle in a Double Slab (Range Modulator)}
Unlike beam spreading in water, the choice of $T$ is important in at least one practical problem in proton transport. The upstream range modulator in a passive beam spreading system is a sequence of high-$Z$/low-$Z$ sandwiches designed to produce a Gaussian fluence distribution of constant width $y_{rms}$ either at the patient or, in a double scattering system, at the second scatterer \cite{delaney,grusell}. Also, each sandwich is designed to pull back the pristine depth-dose by some fixed water equivalent amount so as to produce the desired spread-out Bragg peak.

The design procedure \cite{pbs} is fairly complicated and the details need not concern us. We simply wish to define a sequence of Pb/Lexan/air stacks that might occur in a practical beam line, transport the beam through each such stack, and see how much difference $T$ makes. We could close the loop with an experiment to see which $T$ is best. 

Table \ref{tbl:modStack} lists the parameters of a typical design. $\theta_\mathrm{Hanson,\,3}$ is the angle obtained by combining the Moli\`ere/Fano/Hanson angles for Pb and Lexan in quadrature. Scattering in air is ignored. More important, the single scattering correction for Lexan is wrong because its MCS angle is calculated for a beam of energy $E_2$ entering the Lexan {\em de novo}, ignoring the fact that some MCS has already taken place. For this and other reasons we do not expect the design and transport calculations to agree  exactly for any choice of $T$.   

\begin{figure}[p]
\centering\includegraphics[width=4.58in,height=3.5in]{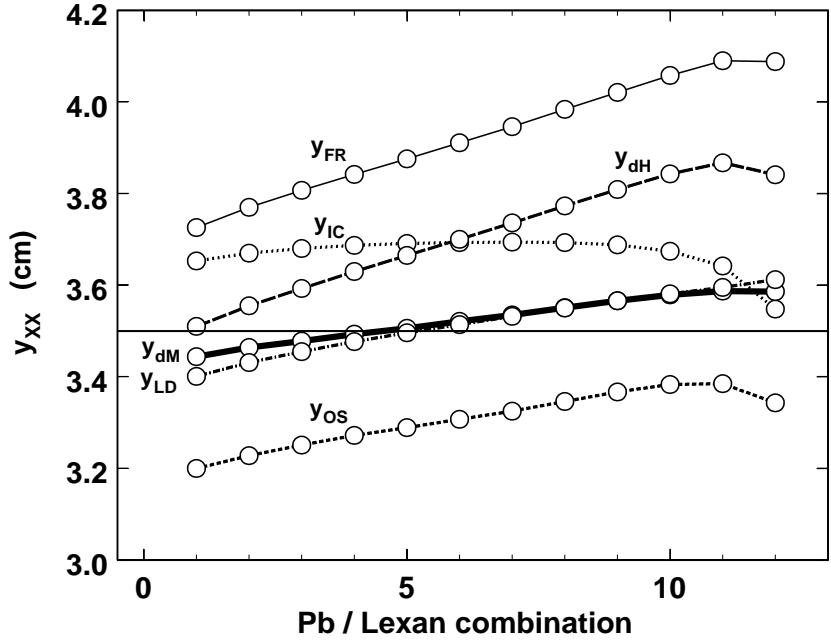} 
\caption{Projected rms displacement $y_\mathrm{XX}$, according to Fermi-Eyges theory with six formulas for $T$, at the end of a sequence of Pb/Lexan/air stacks corresponding to a simplified range modulator (Table\,\ref{tbl:modStack}). The horizontal line at $y_\mathrm{XX}=3.5$\,cm represents the design goal. \label{fig:modStack}}
\end{figure}

Figure \ref{fig:modStack} shows Fermi-Eyges transport results for each Pb/Lexan/air stack, with an ideal incident beam.\footnote{~A known initial beam phase ellipse could easily be used if necessary.} We used Kanematsu's finite increment form (\cite{kanematsu08} Eqs.\,19-21), equivalent to integrating by the midpoint rule. The finite depth of the midpoint of the first step sidesteps the divergence of non-local $T$'s at $x=0$. We discretize each slab separately with a single minimum step size parameter $\rho\Delta x=0.1$\;g/cm$^2$ for all materials, yielding typically tens of steps in Pb, a few hundred in Lexan, and one step for air. Increasing $\rho\Delta x$ even by a factor 20 changes $y_{rms}$ less than a percent.

We have not yet done an experiment to see what $y_{rms}$ these or similar setups actually produce, but Figure\;\ref{fig:modStack} is encouraging. The fact that our standard design procedure in fact produces a flat dose at isocenter (matches the second scatterer design) suggests that $y_{rms}=3.5$\,cm is probably correct to a few percent. If, lacking a direct measurement of $y_{rms}$, we assume that to be the case, we find that $T_\mathrm{dM}$ indeed comes closest to the right answer whereas some of the other $T$'s are off by an amount much larger than the 1-2\% a careful experiment could measure.

It is amusing that $T_\mathrm{LD}$ also gives a very good result, even in a non tissue-like problem. That is consistent with Figure\,\ref{fig:ThLD}. $\theta_{rms}$ for a moderately thick Pb slab is slightly low, whereas for a thick plastic slab it would be slightly high. It augurs well for `corrected Rossi' $T$'s in general \cite{soukop} even though the excellent result here may be somewhat accidental (or, depending on the outcome of a measurement, possibly wrong).

One could argue that, because we have an adequate `slabwise-Hanson' design procedure, we again don't really need either beam transport or $T$. And indeed we don't, if we are merely designing beam spreading systems. However, the Pb/Lexan/air stack could be part of a larger Monte Carlo computation \cite{paganetti}. That brings us to the last topic.

\subsection{Monte Carlo Calculations}
Monte Carlo calculations are the gold standard in radiotherapy. A condensed history Monte Carlo should embody a differential model of multiple scattering analogous to scattering power, even if it is not called that. In each (finite but small) step of known material, one needs to compute the increase in the width parameter of some distribution (Gaussian, Moli\`ere or other) from which a random \begin{table}[h]
\setlength{\tabcolsep}{5pt}
\begin{center}\begin{tabular}{rrrr@{\hspace{37pt}}}
\multicolumn{1}{c}{STEPMAX}&           
\multicolumn{1}{c}{\# steps}&
\multicolumn{1}{c}{$\theta_{rms}$}&           
\multicolumn{1}{c}{$(\theta_{rms}/\theta_\mathrm{Hanson}-1)$}\\           
&&
\multicolumn{1}{c}{millirad}&
\multicolumn{1}{c}{\%}\\           
~\\
default&12.1&109.92&-1.1\\
$0.1\times$\,t&19.1&111.13&0.0\\
$0.01\times$\,t&101.6&118.35&6.5\\
$0.001\times$\,t&1015.0&123.19&10.9
\end{tabular}\end{center}
\caption{A test using Geant4 v9.1, courtesy L. Urban, CERN. 158.6\,MeV protons scatter in 20.196 g/cm$^2$ Pb ($\theta_{rms,\mathrm{expt}}=108\pm1$\,mrad, $\theta_\mathrm{Hanson}=111.1$\,mrad). STEPMAX governs the step size and $\theta_{rms}$ is obtained by fitting the projected angular distribution with a Gaussian.\label{tbl:Geant4}}
\end{table}
deflection angle is then drawn. Just as in deterministic calculations, one would like the final result to be independent of step size over some reasonable range. Table \ref{tbl:Geant4} shows for a simple case that results from Geant\,4, whose MCS model is based on a variant of Highland's formula \cite{geant4}, depend somewhat on step size. Monte Carlos based on a `corrected Rossi' formula \cite{soukop}, similar to $T_\mathrm{LD}$, should not have this problem, but may not always give the right answer (Figure\,\ref{fig:ThLD}) since they lack the single scattering correction. $T_\mathrm{dM}$, easily generalized to other charged particles, might be a good compromise between accuracy and step size independence.

\section{Summary}\label{sec:summary}
Unlike stopping theory, accurate theories of multiple Coulomb scattering such as Moli\`ere's do not flow from a differential form. If that is needed, for deterministic or Monte Carlo transport calculations, it must be devised retroactively as an approximation to the more exact theory. To be accurate for thin scatterers it must include a single scattering correction. That implies nonlocality: in addition to energy and properties of the material at the POI, it must depend in some way on how far the multiple scattering process has advanced. That does not conflict with being numerically integrable in the usual sense that an approximating sum approaches a limit as the step size decreases. 

Nonlocality may be characterized in different ways. Schneider et al. use a generalized definition of the normalized depth of the POI, Kanematsu uses a radiative pathlength integral up to the POI, and we use the diminution of $pv$ from its incident value to the POI. All three can be tested against Moli\`ere theory for uniform slabs, and all three generalize to mixed slabs, as they must to be useful.

We have reviewed three local and two nonlocal formulas for $T$, comparing $\theta_{rms}$ from each one graphically with the `correct' MCS angle $\theta_\mathrm{Hanson}$. One of them, $T_{IC}$, can be simplified for protons by introducing a new material parameter, the scattering length $X_S$, which is similar in form to radiation length $X_0$. $T_\mathrm{IC}$ provides a basis for a new nonlocal scattering power $T_\mathrm{dM}$ (Eqs.\;\ref{eqn:XS},\ref{eqn:XSBragg},\ref{eqn:TdM} and \ref{eqn:fdM}) which, for single slabs with $0.001\le x/R_1\le0.97$, reproduces $\theta_\mathrm{Hanson}$ as well as the `correct' scattering power $T_\mathrm{Hanson}$ to a few percent.

In practical problems, the choice of $T$ is frequently not critical. We have shown, for instance, that all $T$'s described here except $T_\mathrm{FR}$ work well for beam spreading in water and, presumably, water-like materials. In particular $T_\mathrm{LD}$, typical of  `corrected Rossi' formulas, works well. Beam spreading in Pb and, presumably, other high-$Z$ materials, is more sensitive to $T$. 

Turning to a mixed slab problem, we have shown that Pb/Lexan/air combinations typical of an upstream range modulator yield very different answers for different $T$'s, so that a direct experimental test should be easy. Again in this case, $T_\mathrm{LD}$ (even though used outside its supposed range of validity) yields almost the same answer as $T_\mathrm{dM}$, probably because of a lucky cancelation.

In the end, perhaps the strongest case for an accurate scattering power such as $T_\mathrm{dM}$ can be made for general purpose Monte Carlo codes which are supposed to do everything well, rather than beam design or patient dose computations where special purpose workarounds can be devised.

\clearpage
\section{Acknowledgments}
We are indebted to Drs. Kanematsu, Schneider and Hollmark for correspondence regarding their work and to Dr. L. Urban of CERN for generating the data for Table\;\ref{tbl:Geant4}. We particularly thank the Harvard Physics Department and the Laboratory for Particle Physics and Cosmology (LPPC)  for ongoing support.

\appendix
\section{Computational Details}\label{sec:computational}
Examples, figures and tables were computed with Fortran programs which may be downloaded free: see \bs BGware.zip at \BGsite. Source code is in \bs BGware\bs source. All calculations are single precision.

Formulas for variants of Moli\`ere theory are given in \cite{mcsbg}, implemented by module THETA0.FOR. The only significant change is that we now use cubic spline interpolation of $\ln R(\ln T)$\footnote{~In our code $T$ is kinetic energy, $E$ is total energy, $R$ is in g/cm$^2$ and the longitudinal coordinate is usually $z$.}, rather than a polynomial fit, to interpolate range-energy tables (RANGE.FOR).\footnote{~In general, we interpolate published tables for range-energy relations but compute MCS quantities directly from one of the applicable theories. We use few analytical approximations, but many lookup tables.} Subroutine TOUT computes the energy out of a stack of slabs, as well as the outgoing rms projected MCS angle using what we have called the slabwise-Hanson or slabwise-Highland procedure, not Fermi-Eyges transport.

Figures and tables specifically for this paper were computed with various branches of SPWR.FOR. ProjScatPower(x) computes the projected scattering power at $x$ in a stack according to the formula selected by software switch spMode. WhatsHere(x...) returns material properties and such quantities as Kanematsu's $\ell$ and Schneider's $t$ at $x$. The body of each table set in {\LaTeX} was produced as a text file to avoid errors of transcription. Data matrices required for graphics were also imported from text files.

We computed isolated examples with our `proton desk calculator' LOOKUP, a WinXP executable distributed with BGware. LOOKUP is a convenient driver for some of the subroutines mentioned, offering a choice of `tasks'. A useful one in the present context is STACK, which computes various quantities as a proton beam proceeds down a stack of slabs.

Results depend somewhat on the choice of range-energy tables. Common tables differ by 1-2\% for a given material, presumably because of different choices of $I$, the mean excitation energy. We used ICRU Report 49 \cite{icru49} throughout (our table MIXED.RET in \bs BGware\bs data) despite some experimental evidence \cite{moyers} that Janni's 1982 tables \cite{janni82} are better for water. Reference \cite{mcsbg} used Janni's 1966 tables \cite{janni66}.

Finally, some comments on integration. We need to evaluate $A_0$ (Eq.\;\ref{eqn:An}) at many values of $x/R_1$ for the numerous graphs of $\theta_\mathrm{XX}$ compared with $\theta_\mathrm{Hanson}$. By far the most efficient way is to divide the slab by equal ratios so that the contributions of each step to the sum are nearly equal instead of very different.\footnote{~There is no advantage in dividing the higher Fermi-Eyges integrals $A_1$ and $A_2$ by ratios.} If, in addition, Simpson's rule is used rather than the midpoint rule, extremely fast and accurate integration is achieved. The math (subroutine SimpRat in module SPWRSUBS.FOR) is slightly confusing. Suppose we wish to divide a slab of thickness $x$ into $n$ steps $(\Delta x)_1\ldots(\Delta x)_n$ such that the ratio of successive steps is a constant $r$. Let us assume some provisional value for $n$. The formula for the sum of a geometric progression
\[1+r+r^2+\cdots+r^{n-1}\;=\;(r^n-1)/(r-1)\]
leads directly to formulas for the last and first steps
\[(\Delta x)_n\;=\;\frac{r-1}{r^n-1}\,x\qquad,\qquad(\Delta x)_1\;=\;r^{n-1}\;(\Delta x)_n\]
and we simply divide by $r$ on each iteration. That works for any integer $n\ge1$ and any $r>1$. It is physically reasonable to let
\[r\;=\;\left(\frac{R_1}{R_1-x}\right)^{1/n}\]  
because the need to subdivide at all (variation of $pv$) occurs only when the residual range $(R_1-x)$ is significantly smaller than  the range $R_1$. By the same argument it is foolish to pick $n$ at the outset, since that leads to unnecessary subdivision of thin slabs. Therefore we input some desired maximum value of $r$ and let
\[n\;=\;\mathrm{INT}\left(1.+\,\frac{\ln(R_1/(R_1-x))}{\ln (r_\mathrm{max})}\right)\]
For $R_1-x=0.03\,R_1$ and $r_\mathrm{max}=1.6$ we find $n=8$, the most steps needed in practice.

Unlike the midpoint rule, Simpson's rule uses $T(0)$ where any nonlocal formula for $T$ diverges because the single scattering correction diverges. That is easily fixed by substituting a very small positive value, a given fraction of the first step size, for 0. Final results are quite insensitive to that fraction, which we have adjusted to suppress discontinuities, as $n$ takes on successive values, in the graphs of $\theta_\mathrm{XX}$ vs. $x/R_1$.  

\section{Preston and Koehler's Formula}\label{sec:PK}
Preston and Koehler \cite{preston} show that, to a good approximation, the rms radius $\sigma=\sqrt{2}\;y_{rms}$ at any normalized depth $t$ in any single slab is related to its maximum value $\sigma_0$ by
\begin{equation}\label{eqn:PK15}
\frac{\sigma}{\sigma_0}\;=\;\left[2(1-t)^2\ln\frac{1}{1-t}\,+\,3t^2\,-\,2t\right]^{1/2}
\end{equation}
where $t\equiv x/R_1$ is the normalized depth. $\sigma_0$ in water is
\begin{equation}\label{eqn:PK12a}
\sigma_0\;=\;0.00627\;F^{1/2}\,R_1\;^{0.964}\quad\mathrm{cm}
\end{equation}
where
\begin{equation}\label{eqn:PK9}
F\;=\;\sum_i\frac{w_iZ_i(Z_i+1)}{A_i}\ln\left\{\frac{106}{\beta}\left[Z_i^{1/3}(Z_i+1)\rho R_1/A_i\right]^{1/2}\right\}
\end{equation}
and $\beta$ is given by (\ref{eqn:betasq}) with $E$ corresponding to the kinetic energy at a depth $R_1/2$. 

Their derivation of (\ref{eqn:PK15}) involves the {\O}ver{\aa}s approximation which they apparently discovered independently. They use a quantity similar to $T_{IC}$ based on the multiple scattering formalism of Bethe and Ashkin \cite{ashkin}. Kanematsu \cite{kanematsuLD} gives a more modern derivation  of (\ref{eqn:PK15}) based on $T_\mathrm{LD}$.

\clearpage
\bibliographystyle{unsrt}
\bibliography{/pctexv4/work/pbs/master}

\begin{thebibliography}{10}

\bibitem{icru49}
M.J. Berger, M. Inokuti, H.H. Andersen, H. Bichsel, D. Powers, S.M. Seltzer, D.
  Thwaites, D.E. Watt, H. Paul and R.M. Sternheimer, `Stopping Powers and
  Ranges for Protons and Alpha Particles,' ICRU Report 49 (1993).

\bibitem{moliere2}
G. Moli\`ere, `Theorie der Streuung schneller geladenen Teilchen \caproman2
  Mehrfach- und Vielfachstreuung,' Z. Naturforschg. {\bf3a} (1948) 78-97.

\bibitem{bethe}
H.A. Bethe, `Moli\`ere's theory of multiple scattering,' Phys. Rev. {\bf89}
  (1953) 1256-1266. Four entries in the second column (the Gaussian) of Table
  \caproman2 are slightly incorrect (A. Cormack, priv. comm.) but the error
  (corrected in our programs) is at worst 1\%.

\bibitem{mcsbg}
B. Gottschalk, A.M. Koehler, R.J. Schneider, J.M. Sisterson and M.S. Wagner,
  `Multiple Coulomb scattering of 160 MeV protons,' Nucl. Instr. Meth. {\bf
  B74} (1993) 467-490. We have discovered the following errors: Eq.(2) should
  read \[\Xi(\chi)=\frac{1}{\pi}\;\frac{\chi_c^2}{(\chi^2+\chi_a^2)^2}\] and in
  Table 1 the heading $\alpha$ should read $\alpha^2$ and $\times10^9$ under
  $\chi_c^2$ should read $\times10^6$.

\bibitem{icru35}
H. Svensson, P. Almond, A. Brahme, A. Dutreix and H.K. Leetz, `Radiation
  Dosimetry: Electron Beams with Energies Between 1 and 50\;MeV,' ICRU Report
  35 (1984).

\bibitem{hollmark}
M. Hollmark, J. Uhrdin, D{\v{z}} Belki{\'{c}}, I. Gudowska and A. Brahme,
  `Influence of multiple scattering and energy loss straggling on the absorbed
  dose distributions of therapeutic light ion beams: \caproman1. Analytical
  pencil beam model,' Phys. Med. Biol. {\bf49} (2004) 3247-3265.

\bibitem{kanematsu08}
N. Kanematsu, `Alternative scattering power for Gaussian beam model of heavy
  charged particles,' Nucl. Instr. and Meth. B {\bf266} (2008) 5056-5062.

\bibitem{li}
X. Allen Li and D.W.O. Rogers, `Electron mass scattering powers: Monte Carlo
  and analytical calculations,' Med. Phys. {\bf22~(5)} (1995) 531-541.

\bibitem{russell}
Kellie R. Russell, Erik Grusell and Anders Montelius, `Dose calculations in
  proton beams: range straggling corrections and energy scaling,' Phys. Med.
  Biol. {\bf40} (1995) 1031-1043.

\bibitem{pedroniPencil}
E. Pedroni, S. Scheib, T. {B\"ohringer}, A. Coray, M. Grossmann, S. Lin and A.
  Lomax, `Experi\-mental characterization and physical modeling of the dose
  distribution of scanned proton pencil beams,' Phys. Med. Biol. {\bf50} (2005)
  541-561.

\bibitem{hollmark08}
M. Hollmark, I. Gudowska, D{\v{z}} Belki{\'{c}}, A. Brahme and N. Sobolevsky,
  `An analytical model for light ion pencil beam dose distributions: multiple
  scattering of primary and secondary ions,' Phys. Med. Biol. {\bf53} (2008)
  3477-3491.

\bibitem{delaney}
B. Gottschalk, `Passive Beam Scattering,' Chapter 5A in ``Proton and Charged
  Particle Radio\-thera\-py,'' ed. T.F Delaney and H.M. Kooy, Lippincott
  Williams and Wilkins (2008).

\bibitem{grusell}
Erik Grusell, Anders Montelius, Anders Brahme, {G\"oran} Rikner and Kellie
  Russell, `A general solution to charged particle beam flattening using an
  optimized dual scattering foil technique, with application to proton therapy
  beams,' Phys. Med. Biol. {\bf39} (1994) 2201-2216.

\bibitem{overas}
H. {\O}ver{\aa}s, `On small angle multiple scattering in confined bodies,' CERN
  Yellow Report {\bf60-18} (1960).

\bibitem{schneider01}
U. Schneider, J. Besserer and P. Pemler, `On small angle multiple Coulomb
  scattering of protons in the Gaussian approximation,' Z. Med. Phys. {\bf11}
  (2001) 110-118.

\bibitem{highland}
V.L. Highland, `Some practical remarks on multiple scattering,' Nucl. Instr.
  Meth. {\bf 129} (1975) 497-499 and Erratum, Nucl. Instr. Meth. {\bf 161}
  (1979) 171.

\bibitem{rossiBook}
Bruno Rossi, ``High-Energy Particles'', Prentice-Hall, New York (1952).

\bibitem{brahme}
A. Brahme, `On the optimal choice of scattering foils for electron therapy,'
  technical report TRITA-EPP-17, Royal Institute of Technology, Stockholm,
  Sweden (1972).

\bibitem{kraft}
G. Kraft, `Tumor therapy with heavy charged particles,' Progress in Particle
  and Nuclear Physics {\bf45} (2000) S473-S544.

\bibitem{barkas}
H.W. Barkas, ``Nuclear Research Emulsions,'' Academic Press, New York and
  London (1963).

\bibitem{kanematsuLD}
N. Kanematsu, `Semi-empirical formulation of multiple scattering for Gaussian
  beam model of heavy charged particles stopping in tissue-like matter,' Phys.
  Med. Biol. (2009).

\bibitem{soukop}
M. Soukop, M. Fippel and M. Alber, `A pencil beam algorithm for intensity
  modulated proton therapy derived from Monte Carlo Simulations,' Phys. Med.
  Biol. {\bf50} (2005) 5089-5104.

\bibitem{preston}
W.M. Preston and A.M. Koehler, `The effects of scattering on small proton
  beams,' unpublished manuscript (1968), Harvard Cyclotron Laboratory. A
  facsimile is available in \bs BGdocs.zip at \BGsite.

\bibitem{pbs}
B. Gottschalk, ``Passive Beam Spreading in Proton Radiation Therapy,''
  unpublished book (2004) available in PDF format at \BGsite.

\bibitem{paganetti}
H. Paganetti, H. Jiang, S.-Y. Lee and H.M. Kooy, `Accurate Monte Carlo
  simulations for nozzle design, commissioning and quality assurance for a
  proton radiation therapy facility,' Med. Phys. {\bf31~(7)} (2004) 2107-2118.

\bibitem{geant4}
`Geant4 Physics Reference Manual,' download from
  http://geant4.cern.ch/G4UsersDocuments/\\UsersGuides/PhysicsReferenceManual/%
html/PhysicsReferenceManual.html (2008).

\bibitem{moyers}
M.F. Moyers, G.B. Coutrakon, A. Ghebremedhin, K. Shahnazi, P. Koss and E.
  Sanders, `Calibration of a proton beam energy monitor,' Med. Phys. {\bf34(6)}
  (2007) 1952-1966.

\bibitem{janni82}
J.F. Janni, `Proton Range-Energy Tables, 1KeV - 10 GeV,' Atomic Data and
  Nuclear Data Tables {\bf 27} parts 1 (compounds) and 2 (elements) (Academic
  Press, 1982).

\bibitem{janni66}
J.F. Janni, `Calculations of energy loss, range, pathlength, straggling,
  multiple scattering, and the probability of inelastic nuclear collisions for
  0.1 to 1000 MeV protons,' Air Force Weapons Laboratory Technical Report No.
  AFWL-TR-65-150 (1966).

\bibitem{ashkin}
H.A. Bethe and J. Ashkin, `Passage of radiations through matter,' in
  Experimental Nuclear Physics, E. Segr\`e (ed.), Wiley, New York (1953): Part
  \caproman2, page 283 et seg.

\end{thebibliography}

\clearpage
\listoftables


\begin{table}[h]
\begin{center}\begin{tabular}{rrrrrrrrrrr}
\multicolumn{6}{l}{158.60\,MeV p on Be, $\rho R_1=21.290$\,g/cm$^2$ :}\\
\\
\multicolumn{1}{c}{$x/R_1$}&           
\multicolumn{1}{c}{$\rho x$}&           
\multicolumn{1}{c}{$E$}&           
\multicolumn{1}{c}{$\theta_\mathrm{Hanson}$}&           
\multicolumn{1}{c}{$\theta_\mathrm{Highland}$}&           
\multicolumn{1}{c}{$\theta_\mathrm{FR}$}&           
\multicolumn{1}{c}{$\theta_\mathrm{IC}$}&           
\multicolumn{1}{c}{$\theta_\mathrm{LD}$}&           
\multicolumn{1}{c}{$\theta_\mathrm{{\O}S}$}&           
\multicolumn{1}{c}{$\theta_\mathrm{dH}$}&           
\multicolumn{1}{c}{$\theta_\mathrm{dM}$}\\           
\multicolumn{1}{c}{}&           
\multicolumn{1}{c}{g/cm$^2$}&           
\multicolumn{1}{c}{MeV}&           
\multicolumn{1}{c}{mrad}&           
\multicolumn{1}{c}{\%}&           
\multicolumn{1}{c}{\%}&           
\multicolumn{1}{c}{\%}&           
\multicolumn{1}{c}{\%}&           
\multicolumn{1}{c}{\%}&           
\multicolumn{1}{c}{\%}&           
\multicolumn{1}{c}{\%}\\           
 0.001&  0.021& 158.51&   0.565& -6.15& 62.86& 36.65& 45.63& 12.10& -6.51& -3.03\\
 0.010&  0.213& 157.69&   2.020& -1.62& 44.55& 21.29& 29.23&  0.59& -1.89& -0.22\\
 0.100&  2.129& 149.32&   7.207&  3.16& 31.43& 10.28& 17.30& -1.52&  3.10&  1.63\\
 0.200&  4.258& 139.62&  10.784&  4.56& 28.12&  7.50& 14.11& -0.43&  4.72&  2.02\\
 0.500& 10.645& 107.00&  19.812&  6.31& 23.93&  3.98&  9.45& -2.17&  7.18&  2.21\\
 0.900& 19.161&  43.71&  37.994&  7.35& 21.34&  1.82&  3.94& -8.96&  9.81&  1.14\\
 0.970& 20.651&  22.49&  48.051&  7.76& 21.33&  1.80&  1.27&-13.65& 11.00&  0.14\\
\\
\multicolumn{6}{l}{158.60\,MeV p on Al, $\rho R_1=22.372$\,g/cm$^2$ :}\\            
\\
\multicolumn{1}{c}{$x/R_1$}&           
\multicolumn{1}{c}{$\rho x$}&           
\multicolumn{1}{c}{$E$}&           
\multicolumn{1}{c}{$\theta_\mathrm{Hanson}$}&           
\multicolumn{1}{c}{$\theta_\mathrm{Highland}$}&           
\multicolumn{1}{c}{$\theta_\mathrm{FR}$}&           
\multicolumn{1}{c}{$\theta_\mathrm{IC}$}&           
\multicolumn{1}{c}{$\theta_\mathrm{LD}$}&           
\multicolumn{1}{c}{$\theta_\mathrm{{\O}S}$}&           
\multicolumn{1}{c}{$\theta_\mathrm{dH}$}&           
\multicolumn{1}{c}{$\theta_\mathrm{dM}$}\\           
\multicolumn{1}{c}{}&           
\multicolumn{1}{c}{g/cm$^2$}&           
\multicolumn{1}{c}{MeV}&           
\multicolumn{1}{c}{mrad}&           
\multicolumn{1}{c}{\%}&           
\multicolumn{1}{c}{\%}&           
\multicolumn{1}{c}{\%}&           
\multicolumn{1}{c}{\%}&           
\multicolumn{1}{c}{\%}&           
\multicolumn{1}{c}{\%}&           
\multicolumn{1}{c}{\%}\\           
 0.001&  0.022& 158.51&   1.004& -3.33& 54.97& 41.63& 38.57& 14.29& -3.67&  0.56\\
 0.010&  0.224& 157.68&   3.658& -1.82& 34.83& 23.22& 20.54&  1.01& -2.08&  1.43\\
 0.100&  2.237& 149.22&  13.214&  0.82& 21.11& 10.69&  8.07&  0.54&  0.77&  2.06\\
 0.200&  4.474& 139.42&  19.836&  1.68& 17.72&  7.59&  4.80&  2.61&  1.83&  2.14\\
 0.500& 11.186& 106.52&  36.631&  2.72& 13.45&  3.68&  0.06&  0.93&  3.53&  1.94\\
 0.900& 20.135&  42.99&  70.843&  3.28& 10.78&  1.25& -5.39& -7.92&  5.56&  0.56\\
 0.970& 21.701&  21.83&  90.068&  3.69& 10.81&  1.27& -7.50&-14.43&  6.70& -0.45\\
\\
\multicolumn{6}{l}{158.60\,MeV p on Cu, $\rho R_1=26.258$\,g/cm$^2$ :}\\              
\\
\multicolumn{1}{c}{$x/R_1$}&           
\multicolumn{1}{c}{$\rho x$}&           
\multicolumn{1}{c}{$E$}&           
\multicolumn{1}{c}{$\theta_\mathrm{Hanson}$}&           
\multicolumn{1}{c}{$\theta_\mathrm{Highland}$}&           
\multicolumn{1}{c}{$\theta_\mathrm{FR}$}&           
\multicolumn{1}{c}{$\theta_\mathrm{IC}$}&           
\multicolumn{1}{c}{$\theta_\mathrm{LD}$}&           
\multicolumn{1}{c}{$\theta_\mathrm{{\O}S}$}&           
\multicolumn{1}{c}{$\theta_\mathrm{dH}$}&           
\multicolumn{1}{c}{$\theta_\mathrm{dM}$}\\           
\multicolumn{1}{c}{}&           
\multicolumn{1}{c}{g/cm$^2$}&           
\multicolumn{1}{c}{MeV}&           
\multicolumn{1}{c}{mrad}&           
\multicolumn{1}{c}{\%}&           
\multicolumn{1}{c}{\%}&           
\multicolumn{1}{c}{\%}&           
\multicolumn{1}{c}{\%}&           
\multicolumn{1}{c}{\%}&           
\multicolumn{1}{c}{\%}&           
\multicolumn{1}{c}{\%}\\           
 0.001&  0.026& 158.51&   1.545& -1.71& 49.12& 39.85& 33.34& 11.88& -1.99& -0.60\\
 0.010&  0.263& 157.67&   5.664& -1.56& 28.90& 20.88& 15.23& -1.66& -1.79& -0.44\\
 0.100&  2.626& 149.14&  20.535&  0.17& 15.40&  8.22&  2.95& -1.80&  0.13& -0.18\\
 0.200&  5.252& 139.25&  30.855&  0.81& 12.10&  5.13& -0.24&  0.46&  0.97& -0.15\\
 0.500& 13.129& 106.08&  57.059&  1.65&  8.02&  1.31& -4.83& -0.95&  2.43& -0.36\\
 0.900& 23.632&  42.22& 110.652&  2.44&  5.85& -0.73& -9.79& -9.29&  4.66& -1.41\\
 0.970& 25.470&  21.14& 140.987&  3.27&  6.32& -0.29&-10.85&-15.57&  6.20& -2.07\\
\\
\multicolumn{6}{l}{158.60\,MeV p on Pb, $\rho R_1=36.057$\,g/cm$^2$ :}\\                
\\
\multicolumn{1}{c}{$x/R_1$}&           
\multicolumn{1}{c}{$\rho x$}&           
\multicolumn{1}{c}{$E$}&           
\multicolumn{1}{c}{$\theta_\mathrm{Hanson}$}&           
\multicolumn{1}{c}{$\theta_\mathrm{Highland}$}&           
\multicolumn{1}{c}{$\theta_\mathrm{FR}$}&           
\multicolumn{1}{c}{$\theta_\mathrm{IC}$}&           
\multicolumn{1}{c}{$\theta_\mathrm{LD}$}&           
\multicolumn{1}{c}{$\theta_\mathrm{{\O}S}$}&           
\multicolumn{1}{c}{$\theta_\mathrm{dH}$}&           
\multicolumn{1}{c}{$\theta_\mathrm{dM}$}\\           
\multicolumn{1}{c}{}&           
\multicolumn{1}{c}{g/cm$^2$}&           
\multicolumn{1}{c}{MeV}&           
\multicolumn{1}{c}{mrad}&           
\multicolumn{1}{c}{\%}&           
\multicolumn{1}{c}{\%}&           
\multicolumn{1}{c}{\%}&           
\multicolumn{1}{c}{\%}&           
\multicolumn{1}{c}{\%}&           
\multicolumn{1}{c}{\%}&           
\multicolumn{1}{c}{\%}\\           
 0.001&  0.036& 158.51&   2.638&  2.55& 45.33& 42.50& 29.95& 10.10&  2.24&  1.39\\
 0.010&  0.361& 157.65&   9.878& -0.31& 23.07& 20.68& 10.02& -5.13& -0.53& -0.51\\
 0.100&  3.606& 148.96&  36.222& -0.35&  8.99&  6.87& -2.80& -5.95& -0.38& -1.35\\
 0.200&  7.211& 138.91&  54.568& -0.10&  5.66&  3.60& -6.05& -3.83&  0.06& -1.53\\
 0.500& 18.029& 105.27& 101.274&  0.41&  1.71& -0.27&-10.62& -5.13&  1.17& -1.86\\
 0.900& 32.452&  40.97& 197.440&  1.56&  0.15& -1.80&-15.15&-12.62&  3.71& -2.51\\
 0.970& 34.976&  19.95& 253.086&  2.92&  1.15& -0.82&-14.93&-18.53&  5.78& -2.74
\end{tabular}\end{center}
\caption{$\theta_{rms}$ according to eight models for various normalized thicknesses and materials. From left: normalized slab thickness, actual slab thickness, outgoing energy, $\theta_\mathrm{Hanson}$ in milliradians. Remaining seven entries are difference from $\theta_\mathrm{Hanson}$ expressed in \%, $100\times(\theta_\mathrm{XX}/\theta_\mathrm{Hanson}-1)$. $\theta_\mathrm{LD}$ is included for reference only; it is not supposed to be valid for non tissue like matter.\label{tbl:theta0Single}}
\end{table}


\begin{table}[h]
\setlength{\tabcolsep}{5pt}
\begin{center}
\begin{tabular}{rrrrrrrrrrr}
\multicolumn{6}{l}{158.60\,MeV p on Be, $\rho R_1=21.290$\,g/cm$^2$ :}\\
\\
\multicolumn{1}{c}{$x/R_1$}&           
\multicolumn{1}{c}{$\rho x$}&           
\multicolumn{1}{c}{$E$}&           
\multicolumn{1}{c}{$(T/\rho)_\mathrm{Hans}$}&           
\multicolumn{1}{c}{Highland}&           
\multicolumn{1}{c}{FR}&           
\multicolumn{1}{c}{IC}&           
\multicolumn{1}{c}{LD}&           
\multicolumn{1}{c}{{\O}S}&           
\multicolumn{1}{c}{dH}&           
\multicolumn{1}{c}{dM}\\           
\multicolumn{1}{c}{}&           
\multicolumn{1}{c}{g/cm$^2$}&           
\multicolumn{1}{c}{MeV}&           
\multicolumn{1}{c}{mr$^2$cm$^2$/g}&           
\multicolumn{1}{c}{\%}&           
\multicolumn{1}{c}{\%}&           
\multicolumn{1}{c}{\%}&           
\multicolumn{1}{c}{\%}&           
\multicolumn{1}{c}{\%}&           
\multicolumn{1}{c}{\%}&           
\multicolumn{1}{c}{\%}\\           
 0.001&  0.021& 158.51&    16.70& -9.04&138.70& 68.05& 90.84& 13.36& -8.42& -2.16\\
 0.010&  0.213& 157.69&    20.99&  0.47& 91.78& 35.02& 53.23& -5.03&  0.60&  1.94\\
 0.100&  2.129& 149.32&    27.65& 10.18& 61.23& 13.51& 27.96& -0.36& 10.81&  4.75\\
 0.200&  4.258& 139.62&    32.89& 12.84& 53.74&  8.24& 21.00& -0.38& 13.85&  4.79\\
 0.500& 10.645& 107.00&    57.19& 15.46& 46.27&  2.98& 11.29& -8.86& 19.07&  4.08\\
 0.900& 19.161&  43.71&   324.65& 16.67& 45.37&  2.35& -2.40&-36.96& 25.45& -1.16\\
 0.970& 20.651&  22.49&  1168.89& 18.80& 49.29&  5.11&-10.78&-38.70& 29.78& -4.89\\
\\
\multicolumn{6}{l}{158.60\,MeV p on Al, $\rho R_1=22.372$\,g/cm$^2$ :}\\            
\\
\multicolumn{1}{c}{$x/R_1$}&           
\multicolumn{1}{c}{$\rho x$}&           
\multicolumn{1}{c}{$E$}&           
\multicolumn{1}{c}{$(T/\rho)_\mathrm{Hans}$}&           
\multicolumn{1}{c}{Highland}&           
\multicolumn{1}{c}{FR}&           
\multicolumn{1}{c}{IC}&           
\multicolumn{1}{c}{LD}&           
\multicolumn{1}{c}{{\O}S}&           
\multicolumn{1}{c}{dH}&           
\multicolumn{1}{c}{dM}\\           
\multicolumn{1}{c}{}&           
\multicolumn{1}{c}{g/cm$^2$}&           
\multicolumn{1}{c}{MeV}&           
\multicolumn{1}{c}{mr$^2$cm$^2$/g}&           
\multicolumn{1}{c}{\%}&           
\multicolumn{1}{c}{\%}&           
\multicolumn{1}{c}{\%}&           
\multicolumn{1}{c}{\%}&           
\multicolumn{1}{c}{\%}&           
\multicolumn{1}{c}{\%}&           
\multicolumn{1}{c}{\%}\\           
 0.001&  0.022& 158.51&    51.08& -6.29&111.92& 77.00& 69.43& 15.67& -5.67&  3.21\\
 0.010&  0.224& 157.68&    66.29& -1.87& 64.90& 37.73& 31.75& -4.44& -1.73&  4.09\\
 0.100&  2.237& 149.22&    89.29&  3.83& 35.73& 13.37&  7.63&  6.25&  4.33&  4.69\\
 0.200&  4.474& 139.42&   106.54&  5.43& 29.21&  7.92&  1.51&  7.56&  6.59&  4.55\\
 0.500& 11.186& 106.52&   187.74&  6.85& 22.01&  1.91& -7.64& -3.46& 10.10&  3.01\\
 0.900& 20.135&  42.99&  1095.53&  7.68& 20.87&  0.95&-19.14&-42.14& 15.29& -2.65\\
 0.970& 21.701&  21.83&  4015.91& 10.48& 25.13&  4.51&-21.49&-45.01& 20.19& -5.72\\
\\
\multicolumn{6}{l}{158.60\,MeV p on Cu, $\rho R_1=26.258$\,g/cm$^2$ :}\\              
\\
\multicolumn{1}{c}{$x/R_1$}&           
\multicolumn{1}{c}{$\rho x$}&           
\multicolumn{1}{c}{$E$}&           
\multicolumn{1}{c}{$(T/\rho)_\mathrm{Hans}$}&           
\multicolumn{1}{c}{Highland}&           
\multicolumn{1}{c}{FR}&           
\multicolumn{1}{c}{IC}&           
\multicolumn{1}{c}{LD}&           
\multicolumn{1}{c}{{\O}S}&           
\multicolumn{1}{c}{dH}&           
\multicolumn{1}{c}{dM}\\           
\multicolumn{1}{c}{}&           
\multicolumn{1}{c}{g/cm$^2$}&           
\multicolumn{1}{c}{MeV}&           
\multicolumn{1}{c}{mr$^2$cm$^2$/g}&           
\multicolumn{1}{c}{\%}&           
\multicolumn{1}{c}{\%}&           
\multicolumn{1}{c}{\%}&           
\multicolumn{1}{c}{\%}&           
\multicolumn{1}{c}{\%}&           
\multicolumn{1}{c}{\%}&           
\multicolumn{1}{c}{\%}\\           
 0.001&  0.026& 158.51&   103.73& -4.34& 94.86& 71.38& 55.79& 10.10& -3.73&  0.04\\
 0.010&  0.263& 157.67&   135.90& -2.17& 50.20& 32.10& 19.99& -9.53& -2.03& -0.07\\
 0.100&  2.626& 149.14&   183.84&  1.93& 23.21&  8.37& -2.36&  2.07&  2.57&  0.14\\
 0.200&  5.252& 139.25&   219.98&  3.18& 17.11&  3.00& -8.12&  3.61&  4.34& -0.16\\
 0.500& 13.129& 106.08&   389.19&  4.57& 10.76& -2.58&-16.49& -6.53&  7.59& -1.51\\
 0.900& 23.632&  42.22&  2292.33&  7.02& 11.71& -1.75&-25.08&-43.41& 14.48& -5.40\\
 0.970& 25.470&  21.14&  8491.42& 12.04& 17.78&  3.59&-19.18&-45.43& 21.51& -6.88\\
\\
\multicolumn{6}{l}{158.60\,MeV p on Pb, $\rho R_1=36.057$\,g/cm$^2$ :}\\                
\\
\multicolumn{1}{c}{$x/R_1$}&           
\multicolumn{1}{c}{$\rho x$}&           
\multicolumn{1}{c}{$E$}&           
\multicolumn{1}{c}{$(T/\rho)_\mathrm{Hans}$}&           
\multicolumn{1}{c}{Highland}&           
\multicolumn{1}{c}{FR}&           
\multicolumn{1}{c}{IC}&           
\multicolumn{1}{c}{LD}&           
\multicolumn{1}{c}{{\O}S}&           
\multicolumn{1}{c}{dH}&           
\multicolumn{1}{c}{dM}\\           
\multicolumn{1}{c}{}&           
\multicolumn{1}{c}{g/cm$^2$}&           
\multicolumn{1}{c}{MeV}&           
\multicolumn{1}{c}{mr$^2$cm$^2$/g}&           
\multicolumn{1}{c}{\%}&           
\multicolumn{1}{c}{\%}&           
\multicolumn{1}{c}{\%}&           
\multicolumn{1}{c}{\%}&           
\multicolumn{1}{c}{\%}&           
\multicolumn{1}{c}{\%}&           
\multicolumn{1}{c}{\%}\\           
 0.001&  0.036& 158.51&   225.07&  1.16& 81.30& 74.31& 44.94&  4.47&  1.78&  1.99\\
 0.010&  0.361& 157.65&   304.43& -1.40& 35.39& 30.17&  8.15&-16.64& -1.25& -1.35\\
 0.100&  3.606& 148.96&   419.99& -0.19&  9.12&  4.91&-13.67& -6.69&  0.30& -2.93\\
 0.200&  7.211& 138.91&   505.18&  0.51&  3.43& -0.56&-19.12& -5.32&  1.43& -3.51\\
 0.500& 18.029& 105.27&   898.09&  1.81& -1.67& -5.46&-26.56&-13.88&  4.72& -4.40\\
 0.900& 32.452&  40.97&  5392.73&  6.85&  1.68& -2.24&-32.37&-46.40& 13.96& -6.13\\
 0.970& 34.976&  19.95& 20673.75& 14.12&  9.48&  5.26&-14.95&-48.30& 23.50& -5.96\\
\end{tabular}
\end{center}
\caption{$(T/\rho)$ according to eight models for various materials and normalized thicknesses. From left: normalized slab thickness, actual slab thickness, outgoing energy, $(T/\rho)_\mathrm{Hanson}$ in milliradian$^2$/(g/cm$^2$). Remaining seven entries are difference from $(T/\rho)_\mathrm{Hanson}$ expressed in \%, $100\times((T/\rho)_\mathrm{XX}/(T/\rho)_\mathrm{Hanson}-1)$. $(T/\rho)_\mathrm{LD}$ is included for reference only; it is not supposed to be valid for non tissue like matter.\label{tbl:ToverRhoSingle}}
\end{table}

\clearpage
\listoffigures

\clearpage

\clearpage

\clearpage
\begin{figure}[p]
\centering\includegraphics[width=4.72in,height=3.5in]{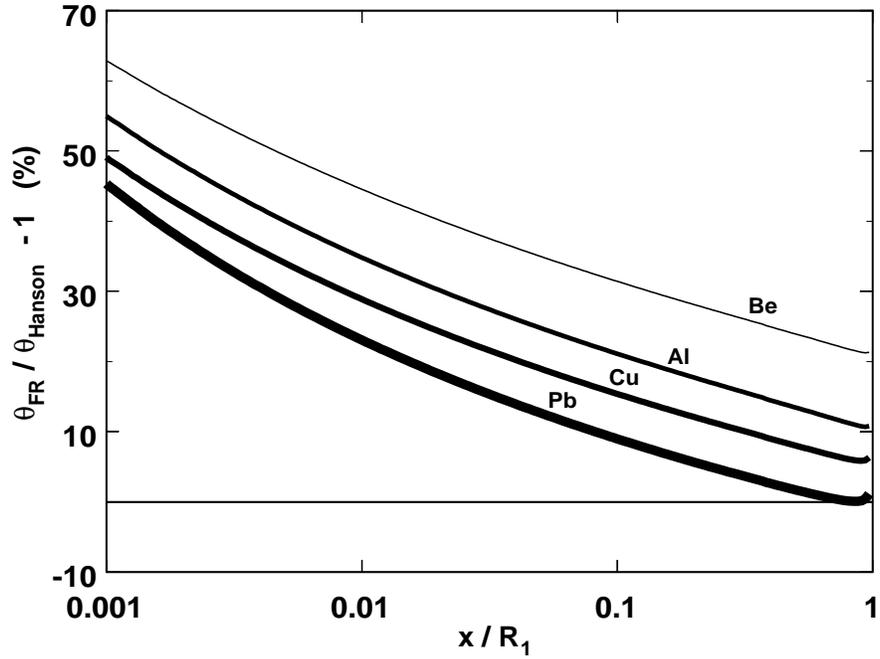} 
\caption{Deviation from $\theta_\mathrm{Hanson}$ of $\theta_\mathrm{FR}$ computed from $T_\mathrm{FR}$, the Fermi-Rossi scattering power, for four scattering materials. The incident proton energy is 158.6\,MeV.
\label{fig:ThR}}
\end{figure}

\clearpage
\begin{figure}[p]
\centering\includegraphics[width=4.72in,height=3.5in]{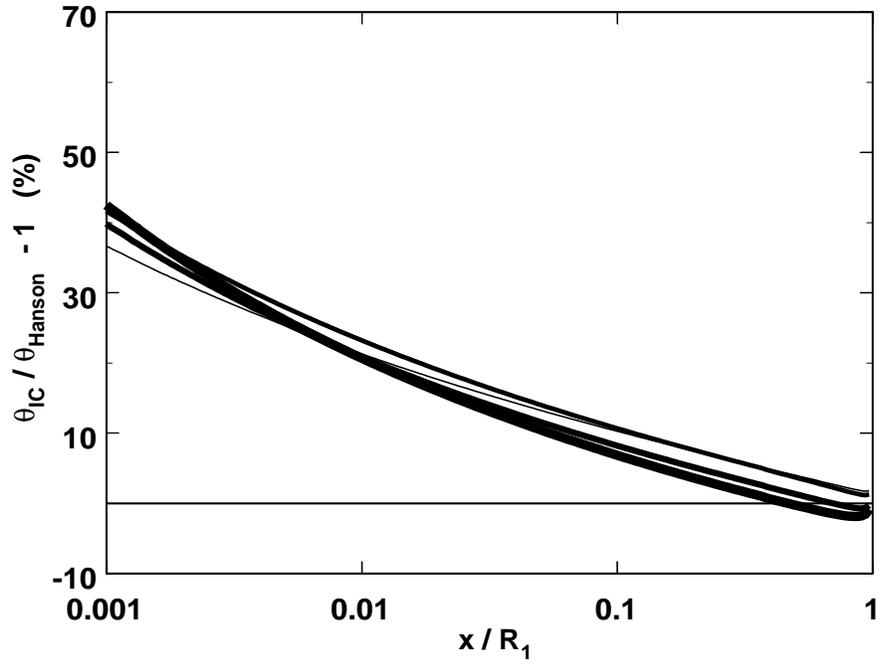} 
\caption{Deviation from $\theta_\mathrm{Hanson}$ of $\theta_\mathrm{IC}$ computed from $T_\mathrm{IC}$, the ICRU35 Report 35-- scattering power adapted to protons, for four scattering materials. The incident proton energy is 158.6\,MeV and the line width code the same as Figure\,\ref{fig:ThR}.
\label{fig:ThI}}
\end{figure}

\clearpage

\clearpage

\clearpage

\clearpage

\clearpage

\clearpage

\clearpage

\begin{table}[p]
\setlength{\tabcolsep}{10pt}
\begin{center}\begin{tabular}{rrrrrrrr}
\multicolumn{1}{c}{comb}&           
\multicolumn{1}{c}{Pb}&
\multicolumn{1}{c}{Lexan}&           
\multicolumn{1}{c}{$E_2$}&           
\multicolumn{1}{c}{$E_3$}&           
\multicolumn{1}{c}{$E_4$}&           
\multicolumn{1}{c}{$\theta_\mathrm{Hanson,\,3}$}&           
\multicolumn{1}{c}{$x_0$}\\           
\multicolumn{1}{c}{\#}&
\multicolumn{1}{c}{g/cm$^2$}&
\multicolumn{1}{c}{g/cm$^2$}&           
\multicolumn{1}{c}{MeV}&           
\multicolumn{1}{c}{MeV}&           
\multicolumn{1}{c}{MeV}&           
\multicolumn{1}{c}{mrad}&           
\multicolumn{1}{c}{cm}\\           
  1&   6.429&   0.000&   216.4&   216.4&   215.9&   35.10&    0.28\\
  2&   6.173&   2.560&   216.9&   206.4&   205.9&   35.12&    0.34\\
  3&   5.872&   5.144&   217.6&   196.1&   195.6&   35.19&    0.53\\
  4&   5.543&   7.743&   218.3&   185.4&   184.9&   35.32&    0.88\\
  5&   5.179&  10.360&   219.1&   174.3&   173.8&   35.51&    1.44\\
  6&   4.781&  12.995&   219.9&   162.7&   162.2&   35.79&    2.23\\
  7&   4.335&  15.656&   220.9&   150.5&   150.0&   36.19&    3.30\\
  8&   3.834&  18.346&   221.9&   137.5&   137.0&   36.73&    4.71\\
  9&   3.240&  21.085&   223.2&   123.7&   123.1&   37.45&    6.55\\
 10&   2.509&  23.898&   224.7&   108.6&   108.0&   38.42&    8.94\\
 11&   1.537&  26.840&   226.8&    91.8&    91.1&   39.78&   12.11\\
 12&   0.000&  30.082&    72.2&    72.2&    71.5&   41.93&   16.54\\
\end{tabular}\end{center}
\caption{Pb/Lexan combinations of the simplified range modulator used in computing Figure \ref{fig:modStack}. $E_i$ is the proton energy entering the $i^\mathrm{th}$ slab ($E_1=230$\,MeV), $\theta_\mathrm{Hanson,\,3}$ is the design MCS angle entering air and $x_0$ is the effective scattering point used in designing the modulator for constant $y_{rms}=3.5$\,cm at 100\,cm. The pullback per position is 2.308\;cm water equivalent.\label{tbl:modStack}}
\end{table}

\end{document}